\documentclass{jfm}

\usepackage[a4paper, total={6.5in, 9.5in}]{geometry}
\usepackage{graphicx}
\usepackage{dcolumn}
\usepackage{bm}
\usepackage{epstopdf, epsfig}

\usepackage{natbib}
\usepackage{xcolor}
\usepackage{amsmath}
\usepackage{lineno}
\usepackage{subfigure}
\usepackage{caption}
\usepackage[normalem]{ulem}
\usepackage{gensymb}

\DeclareFontFamily{OT1}{pzc}{}
\DeclareFontShape{OT1}{pzc}{m}{it}{<-> s * [1.10] pzcmi7t}{}
\DeclareMathAlphabet{\mathpzc}{OT1}{pzc}{m}{it}

\ifCUPmtlplainloaded \else
  \checkfont{eurm10}
  \iffontfound
    \IfFileExists{upmath.sty}
      {\typeout{^^JFound AMS Euler Roman fonts on the system,
                   using the 'upmath' package.^^J}%
       \usepackage{upmath}}
      {\typeout{^^JFound AMS Euler Roman fonts on the system, but you
                   dont seem to have the}%
       \typeout{'upmath' package installed. JFM.cls can take advantage
                 of these fonts,^^Jif you use 'upmath' package.^^J}%
      }
  \else
  \fi
\fi

\ifCUPmtlplainloaded \else
  \checkfont{msam10}
  \iffontfound
    \IfFileExists{amssymb.sty}
      {\typeout{^^JFound AMS Symbol fonts on the system, using the
                'amssymb' package.^^J}%
       \usepackage{amssymb}%

      }{}
  \fi
\fi

\ifCUPmtlplainloaded \else
  \IfFileExists{amsbsy.sty}
    {\typeout{^^JFound the 'amsbsy' package on the system, using it.^^J}%
     \usepackage{amsbsy}}
    {\providecommand\boldsymbol[1]{\mbox{\boldmath $##1$}}}
\fi

\newcommand\Pe{\mbox{\textit{Pe}}}  

\newsavebox{\astrutbox}
\sbox{\astrutbox}{\rule[-5pt]{0pt}{20pt}}

\numberwithin{equation}{section}

\def\XXint#1#2#3{{\setbox0=\hbox{$#1{#2#3}{\int}$}
     \vcenter{\hbox{$#2#3$}}\kern-.5\wd0}}

\newcommand{\rs}{r^{*}}
\newcommand{\zs}{z^{*}}
\newcommand{\ts}{t^{*}}
\newcommand{\hs}{h^{*}}
\newcommand{\rhos}{\rho^{*}}
\newcommand{\mus}{\mu^{*}}
\newcommand{\sigmas}{\sigma^{*}}
\newcommand{\Vs}{V^{*}}
\newcommand{\Rs}{R^{*}}
\newcommand{\us}{u^{*}}

\newcommand{\ps}{p^{*}}
\newcommand{\Es}{E^{*}}
\newcommand{\gs}{g^{*}}

\newcommand{\tit}{\textit}
\newcommand{\tbf}{\textbf}
\newcommand{\mbf}{\mathbf}

\newcommand{\diff}{\mbox{d}}

\newcommand{\Ca}{\mbox{Ca}}
\newcommand{\Bo}{\mbox{Bo}}
\renewcommand{\Pe}{\mbox{Pe}}

\renewcommand{\th}{\tilde{h}}
\newcommand{\tM}{\tilde{\mathcal{M}}}
\newcommand{\tu}{\tilde{u}}
\newcommand{\tr}{\tilde{r}}

\begin{document}

\title[Gravity can lead to multiple peaks in the early stages of coffee ring formation]{Gravity can lead to multiple peaks in the early stages of coffee ring formation}

\author[M. R. Moore \& A. W. Wray]%
{M.\ns R.\ns M\ls O\ls O\ls R\ls E$^1$ \ns 
\and
A.\ns W. \ns W\ls R\ls A\ls Y$^2$}

\affiliation{$^1$Department of Mathematics, School of Natural Sciences, University of Hull, Cottingham Road, Hull, HU6 7RX, UK
\\
$^2$Department of Mathematics and Statistics, University of Strathclyde, Livingstone Tower,
26 Richmond Street, Glasgow G1 1XH, UK}

\pubyear{}
\volume{}
\pagerange{}
\date{?; revised ?; accepted ?. - To be entered by editorial office}

\maketitle


\begin{abstract}
We consider the role of gravity in solute transport when a thin droplet evaporates. Under the physically-relevant assumptions that the contact line is pinned and the solutal P\'{e}clet number, $\Pe$ is large, we identify two fundamental regimes that depend on the size of the Bond number, $\Bo$. When $\Bo = O(1)$, the asymptotic structure of solute transport follows directly from the surface tension-dominated regime, whereby advection drives solute towards the contact line, only to be countered by local diffusive effects, leading to the formation of the famous ``coffee ring". For larger Bond numbers, we identify the distinguished limit in which $\Bo^{-1/2}\Pe^{2/3} = O(1)$, where the diffusive boundary layer is comparable to the surface tension boundary layer. In each regime, we perform a systematic asymptotic analysis of the solute transport and compare our predictions to numerical simulations of the full model. Our analysis identifies the effect of gravity on the nascent coffee ring, providing quantitative predictions of the size, location and shape of the solute mass profile. Furthermore, we reveal that, for certain values of $\Bo$, $\Pe$ and the evaporation time, a secondary peak may exist inside the classical coffee ring. We find that the onset of this secondary peak is linked to the change in behaviour of the critical point in the droplet centre. Both the onset and the peak characteristics are shown to be independent of $\Pe$, but solutal diffusion may act to remove the secondary peak when the classical coffee ring becomes so large as to subsume it.
\end{abstract}

\begin{keywords}
 
\end{keywords}

\section{Introduction}
The evaporation of sessile droplets has received significant attention in recent years, being the subject of several major reviews \citep{cazabat2010evaporation,lohse2015surface,brutin2018recent,wilson2023evaporation} due to its ubiquity in theoretical, experimental and industrial settings. A particular phenomenon of interest is the so-called ``coffee ring effect", in which a solute in such an evaporating droplet ends up preferentially accumulated at the contact line \citep{Deegan1997,Deegan2000}. This effect is very robust, occurring even when the solution is initially uniformly dispersed throughout the droplet, and even when the evaporative flux is not preferentially localised at the contact line \citep{boulogne2016coffee}.

Motivated by typical physical parameters, models of such systems typically assume that the P\'eclet number is sufficiently large that diffusive effects can be neglected, and so dynamics of the solute inside the droplet are governed purely by convection \citep{Deegan1997,Wray2021}. This unphysical assumption leads to a variety of undesirable side-effects, including singular accumulations of residue, and solute not being conserved \citep{Deegan2000}. 

A variety of attempts have been made to resolve this problem phenomenologically, including via the incorporation of jamming effects \citep{popov2005evaporative,kaplan2015evaporation}. However, jamming effects only become significant close to the particle packing fraction, and the assumptions underpinning the model fail long before this point. 
In particular, the assumption that diffusive effects can be ignored breaks down in a diffusive boundary layer close to the contact line \citep{Moore2021a}, as might be anticipated from the singular accumulation in the na\"ive, convection-only model. This boundary layer and its growth and dynamics have been analysed and understood via matched asymptotics and careful numerics in situations where droplets are small, and thus exist at quasi-static equilibrium due to surface tension \citep{Moore2022}, but little is known for larger droplets where the effects of gravity are important.

Investigations of larger droplets have a long history, dating back to numerical integration of the appropriate Laplace equations by \citet{padday1971profiles} and \citet{boucher1975pendent}, with a variety of studies via asymptotics of their shape \citep{rienstra1990shape,o1991shape,yariv2022shape} and stability \citep{pozrikidis2012stability} in the intervening time. The effect of gravity on droplets, and especially their internal flows, has experienced a recent resurgence of interest due to the experiments of \citet{edwards2018density}, which showed that the dynamics of binary droplets can be sensitively dependent on droplet inclination (and hence gravity). This has since received extensive investigation both experimentally and numerically \citep{li2019gravitational,pradhan2017evaporation}.



Notably, however, despite the original experiments of \citet{Deegan1997} involving large droplets,  there have been relatively few investigations of particle transport inside them, with those available being principally experimental \citep{sandu2011influence,hampton2012influence,devlin2016importance}. This is perhaps because of the robustness of the coffee-stain effect: asymptotic and numerical investigations \citep{barash2009evaporation,kolegov2014mathematical} confirm the experimental results that the ring-stain is preserved unless additional physics are incorporated, such as continuous particle deposition \citep{devlin2016importance}. However, this neglects the bulk of the story, including the dynamics of the residue over the course of the lifetime of the droplets: a critical omission in situations such as continuous particle deposition. We show in the present work that the dynamics are actually quite subtle and complex, and certainly merit detailed investigation.

The structure of this paper is therefore as follows. In \textsection \ref{sec:Config}, we describe the equations governing the fluid flow and solute transport for the problem of a thin droplet evaporating under a diffusive flux, in particular highlighting the effect of gravity in the model. We nondimensionalise the model and introduce the three key dimensionless numbers in the model: the capillary, Bond and P\'{e}clet numbers. In \textsection \ref{sec:Flow_solution}, we completely solve for the liquid flow in the limit in which the solute is dilute, so that the flow and solute transport problems decouple. We discuss pertinent features of the resulting fluid velocity and droplet shape, and in particular how these features vary with the Bond number. 

The bulk of the analysis in this paper concerns the influence of gravity on  solute transport within the droplet, which we analyse in the physically-relevant large-P\'{e}clet number limit in \textsection \ref{sec:Asymptotic_Analysis_alpha_O1}. We find that there are two distinct regimes depending on the relative sizes of the Bond and P\'{e}clet numbers. In the first, where the Bond number is moderate, we extend the asymptotic analysis of \cite{Moore2021a} to include the effect of gravity. However, when the Bond number is also large, a more complex asymptotic analysis is necessary, which is presented in detail in Appendix \ref{appendix:Large_Bo}. In each asymptotic regime, we derive predictions for the distribution of the solute mass within the droplet and compare the results to numerical simulations of the full advection-diffusion problem. In particular, while we find the expected `nascent coffee ring' profile in the solute mass, for certain input parameters, we also find evidence of a novel phenomenon whereby a \tit{second} peak may also develop in the mass profile inside the classical coffee ring.



We analyse both of these peaks in detail in \textsection \ref{sec:Peak_analysis}. In particular, for the classical coffee ring, we discuss the effect of gravity in each of the two asymptotic regimes discussed in \textsection \ref{sec:Asymptotic_Analysis_alpha_O1} and Appendix \ref{appendix:Large_Bo}, while for the secondary peak, we investigate the key role gravity plays in its existence and how the secondary peak may also be subsumed in the classical coffee ring for certain values of the Bond and P\'{e}clet numbers. Finally, in \textsection \ref{sec:Summary}, we summarize our findings and discuss implications to various applications, as well as avenues for future study.  

\section{Problem configuration}
\label{sec:Config}

\begin{figure}
\centering \scalebox{0.85}{\epsfig{file=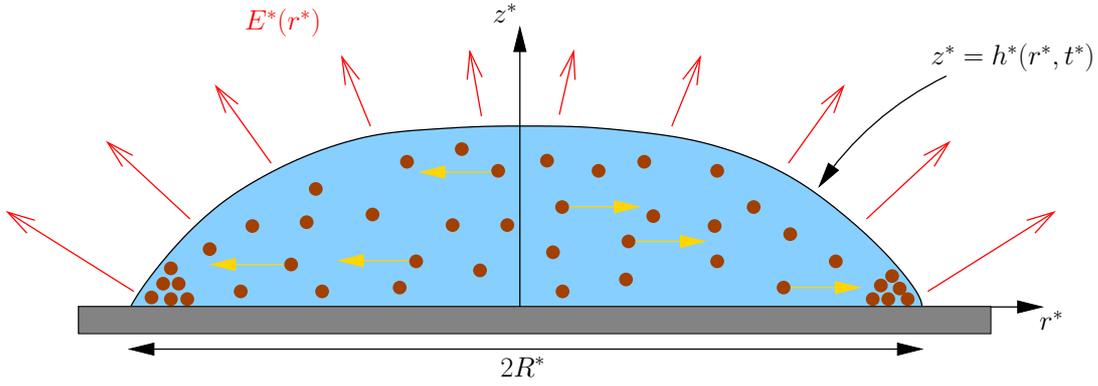}}
\caption{A side-on view of a solute-laden droplet evaporating under an evaporative flux $E^*(r^*)$ from a solid substrate that lies in the plane $z^* = 0$. The droplet is axisymmetric and the contact line is assumed to be pinned on the substrate at $r^* = R^*$. The droplet free surface is denoted by $h^*(r^*,t^*)$. The solute is assumed to be inert and sufficiently dilute that the flow of liquid in the droplet is decoupled from the solute transport.}
\label{fig:Problem_Config} 
\end{figure}

We consider the configuration depicted in figure \ref{fig:Problem_Config} in which an axisymmetric droplet of initial volume $\Vs_{0}$ evaporates from a solid substrate. Here and hereafter, an asterisk denotes a dimensional variable. We let $(\rs,\theta,\zs)$ be cylindrical polar coordinates centred along the line of symmetry of the droplet with the substrate lying in the plane $\zs = 0$: by axisymmetry, we shall assume that all the variables are independent of $\theta$. The droplet contact line is thus circular and we assume that it is \tit{pinned} throughout the drying process, which is observed in practice for a wide range of liquids for the majority of the drying time \citep{Deegan1997,Hu2002,Kajiya2008,howard2023surfactant}. We let $\rs = \Rs$ be the radius of the contact line. Throughout this analysis, we shall assume that the droplet is \tit{thin}, which reduces to the assumption that
\begin{linenomath}\begin{equation}
 0<\delta = \frac{\Vs_{0}}{R^{*3}} \ll1.
\end{equation}\end{linenomath}
As we discuss presently, the thin-droplet assumption allows us to greatly simplify the flow and solute transport models; the assumption has been extensively-validated and has shown to be reasonable even for droplets that should realistically fall outside of this regime \citep{larsson2022quantitative}. 

The droplet consists of a liquid of constant density and viscosity denoted by $\rhos$ and $\mus$, respectively. The droplet free surface is denoted by $\zs = h(\rs,\ts)$ and the air-water surface tension coefficient, $\sigma^*$ is assumed to be constant. 


The liquid evaporates into the surrounding air and we assume that the evaporative process is quasi-steady, which is a reasonable assumption for a wide range of liquid-substrate configurations \citep{Hu2002}. While there are a number of different viable evaporation models depending on the physical and chemical characteristics of the problem \citep{Murisic2011}, for the purposes of this analysis, we assume that the dominant process of vapour transport from the droplet surface is diffusion, so that the evaporative flux $\Es(\rs)$ is given by
\begin{linenomath}
\begin{equation}
 \Es(\rs) = \frac{2D^{*}(c_{s}^{*}-c_{\infty}^{*})}{\pi\sqrt{R^{*2}-r^{*2}}},
 \label{eqn:J}
\end{equation}
\end{linenomath}
where $D^{*}$ is the diffusion coefficient and $c_{s}^{*}$, $c_{\infty}^{*}$ are the surface and ambient vapour concentrations, respectively \citep{Deegan2000, Murisic2011}.

The droplet contains an inert solute of initially uniform concentration $\phi^{*}_{0}$. The solute is assumed to be sufficiently dilute that the flow and transport problems completely decouple. We shall discuss the validity of the dilute assumption further in \textsection \ref{sec:Summary}.

\subsection{Flow model}

The droplet is assumed to be sufficiently thin and the evaporation-induced flow sufficiently slow that the flow is governed by the lubrication equations 
\begin{linenomath}
\begin{alignat}{2}
 \frac{\partial \hs}{\partial \ts} + \frac{1}{\rs}\frac{\partial}{\partial\rs}\left(\rs\hs\us\right) & \; = && \; - \frac{\Es}{\rhos}, \label{eqn:Dim_Lubrication_1}\\
 \us & \; = && \; - \frac{h^{*2}}{3\mus}\frac{\partial\ps}{\partial \rs}, \label{eqn:Dim_Lubrication_2}\\
 \ps & \; = && \; p_{\mathrm{atm}}^{*}-\rhos\gs(\zs-\hs) - \sigmas\frac{1}{\rs}\frac{\partial}{\partial\rs}\left(\rs\frac{\partial\hs}{\partial\rs}\right), \label{eqn:Dim_Lubrication_3} 
\end{alignat}
\end{linenomath}
for $0<\rs<\Rs$, $\ts>0$, where $\us(\rs,\ts)$ is the depth-averaged radial fluid velocity, $\ps(\rs,\zs,\ts)$ is the liquid pressure and $\ps_{\mathrm{atm}}$ denotes atmospheric pressure \citep{hocking1983spreading,Deegan2000,Oliver2015}.

Equations (\ref{eqn:Dim_Lubrication_1})--(\ref{eqn:Dim_Lubrication_3}) must be solved subject to the symmetry conditions
\refstepcounter{equation}
\begin{linenomath}
$$
\rs\hs\us = \frac{\partial\hs}{\partial\rs} = 0 \quad \mbox{at} \quad \rs = 0,
\eqno{(\theequation{\mathit{a},\mathit{b}})}
 \label{eqn:Dim_Symmetry_1}
$$
\end{linenomath}
and the fact that the free surface touches down at, and we require no-flux of liquid through, the pinned contact line, that is
\refstepcounter{equation}
\begin{linenomath}
$$
 \hs = \rs\hs\us =  0 \quad \mbox{at} \quad \rs = \Rs.
   \eqno{(\theequation{\mathit{a},\mathit{b}})}
\label{eqn:Dim_BCs_1}
$$
\end{linenomath}
We close the problem by specifying the initial droplet profile, that is
\begin{linenomath}\begin{equation}
 \hs(\rs,0) = \hs_{0}(\rs) \quad \mbox{for} \quad 0<\rs<\Rs.
 \label{eqn:Dim_ICs_1}
\end{equation}\end{linenomath}

It is worth noting at this stage that, while this initial condition is needed to fully specify the mathematical problem, in our analysis, we do not explicitly use the initial condition (\ref{eqn:Dim_ICs_1}). In what follows, it is assumed that the rate of evaporation is sufficiently slow that the droplet quickly relaxes under capillary action to the quasi-steady profile found in \textsection \ref{sec:Flow} (see, for example, \cite{Lacey1982,deGennes1985,Oliver2015}). Thus, we shall for simplicity assume that $h_{0}(r)$ is of the same functional form of the free surface we find in \textsection \ref{sec:Flow}. While this assumption is reasonable for a wide range of applications, for extremely rapid evaporation (for example, laser-induced evaporation, \cite{volkov2019measuring}), a more careful consideration of the evolution after deposition would be needed.




Assuming the contact line is pinned, the volume of the droplet $\Vs(\ts)$ is given by
\begin{linenomath}\begin{equation}
 \Vs(\ts) = 2\pi\int_{0}^{\Rs}\rs\hs(\rs,\ts)\,\mbox{d}\rs, \quad \Vs(0) = \Vs_{0}.
 \label{eqn:Dim_Volume_1}
\end{equation}\end{linenomath}
The total mass loss due to evaporation $F^{*}(\ts)$ is given by
\begin{linenomath}\begin{equation}
 F^{*}(\ts) = 2\pi\int_{0}^{\Rs}\rs \Es(\rs)\,\mbox{d}\rs = 4D^{*}(c_{s}^{*}-c_{\infty}^{*})\Rs.
\end{equation}\end{linenomath}
Thus, conservation of mass in the liquid phase is
\begin{linenomath}\begin{equation}
 \frac{\diff\Vs}{\diff \ts} = -\frac{F^{*}}{\rho^*} = -\frac{4D^{*}(c_{s}^{*}-c_{\infty}^{*})\Rs}{\rho^*}
\end{equation}\end{linenomath}
so that
\begin{linenomath}\begin{equation}
 \Vs(\ts) = \Vs_{0} - \frac{4D^{*}(c_{s}^{*}-c_{\infty}^{*})\Rs\ts}{\rho^*}.
 \label{eqn:Dim_Volume_2}
\end{equation}\end{linenomath}
In particular, the \tit{dryout time}, that is the time when the drop has fully evaporated, is
\begin{linenomath}\begin{equation}
 t_{f}^{*} = \frac{\rho^*\Vs_{0}}{4D^{*}(c_{s}^{*}-c_{\infty}^{*})\Rs}.
 \label{eqn:Dim_Dryout_Time}
\end{equation}\end{linenomath}


\subsection{Solute model}

The droplet is assumed to be sufficiently thin that the transport of the solute is governed by the depth-averaged advection-diffusion equation 
\begin{linenomath}\begin{equation}
 \frac{\partial}{\partial \ts}\left(\hs\phi^{*}\right) + \frac{1}{\rs}\frac{\partial}{\partial \rs}\left[\rs\left(\hs\us \phi^* - D^{*}_{\phi}\hs\frac{\partial\phi^{*}}{\partial\rs}\right)\right] = 0
 \label{eqn:Dim_Adv_Diff}
\end{equation}\end{linenomath}
for $0<\rs<\Rs$, $\ts>0$, where $\phi^{*}(\rs,\ts)$ is the depth-averaged solute concentration and $D^{*}_{\phi}$ is the solutal diffusion coefficient \citep{Wray2014, Pham2017, Moore2021a}. 

While there is an acknowledged effect of the solute particles eventually being trapped at and transported along the free surface \citep{Kang2016,d2022evolution}, this effect is less pronounced for thin droplets, where the capture tends to occur closer to the contact line due to the stronger outward radial flow. Thus, we shall neglect its effects here as our study concerns the interplay between gravity, surface tension and solute advection/diffusion. A more focused analysis on the final deposit profile would certainly need to account for such effects.

Equation \eqref{eqn:Dim_Adv_Diff} must be solved subject to the symmetry condition
\begin{linenomath}\begin{equation}
 \frac{\partial\phi^{*}}{\partial \rs} = 0 \quad \mbox{at} \quad \rs = 0,
 \label{eqn:Dim_Symmetry_2}
\end{equation}\end{linenomath}
and the condition that there can be no flux of solute particles through the pinned contact line,
\begin{linenomath}\begin{equation}
 \rs\left(\hs\us\phi^* - D^{*}_{\phi}\hs\frac{\partial\phi^{*}}{\partial\rs}\right) = 0 \quad \mbox{at} \quad \rs = \Rs.
 \label{eqn:Dim_BCs_2}
\end{equation}\end{linenomath}
Finally, we impose the initially uniform distribution of solute throughout the droplet, so that
\begin{linenomath}\begin{equation}
 \phi^{*}(\rs,0) = \phi^{*}_{0} \quad \mbox{for} \quad 0<\rs<\Rs.
 \label{eqn:Dim_ICs_2}
\end{equation}\end{linenomath}

\subsection{Non-dimensionalization}

We assume that the fluid velocity is driven by evaporation and, for now, we retain both gravity and surface tension, so that the pertinent scalings are 
\begin{linenomath}\begin{equation}\begin{aligned}
 & (r^{*}, z^{*}) =  \; R^{*}( r,\delta z), \quad u^{*} = 
\frac{D^{*}(c_{s}^{*}-c_{\infty}^{*})}{\delta\rho^{*}\Rs}u, \quad t^{*} = t_{f}^{*}t, \quad   \phi^{*} = \phi^{*}_{0}\phi, \\
  & (h^{*},h_{0}^{*}) = \delta R^{*} (h,h_{0}), \quad p^{*} = p_{\mathrm{atm}}^{*} + \frac{\mu^{*}D^{*}(c_{s}^{*}-c_{\infty}^{*})}{\delta^{3}\rho^{*} R^{*2}} p \quad V^{*} = V_{0}^{*}V.
  \label{eqn:Scalings}
\end{aligned}\end{equation}\end{linenomath}
Note, in particular, that the choice of timescale fixes the dimensionless dryout time to be $t = 1$.

Upon substituting the scalings (\ref{eqn:Scalings}) into (\ref{eqn:Dim_Lubrication_1})--(\ref{eqn:Dim_Lubrication_3}), we see that
\begin{linenomath}
\begin{alignat}{2}
 \frac{\partial h}{\partial t} + \frac{1}{4r}\frac{\partial}{\partial r}\left(rhu\right) & \; = && \; - \frac{1}{2\pi\sqrt{1-r^{2}}}, \label{eqn:NonDim_Lubrication_1}\\
 u & \; = && \; \frac{h^{2}}{3\Ca}\frac{\partial}{\partial r}\left[-\Bo h + \frac{1}{r}\frac{\partial}{\partial r}\left(r\frac{\partial h}{\partial r}\right)\right], \label{eqn:NonDim_Lubrication_2} 
\end{alignat}
\end{linenomath}
for $0<r<1$, $0<t<1$, where the Capillary and Bond numbers are defined by
\begin{linenomath}\begin{equation}
 \Ca = \frac{\mu^{*}D^{*}(c_{s}^{*}-c_{\infty}^{*})}{\delta^{4}\rho^{*}R^{*}\sigma^{*}} \quad \mbox{and} \quad \Bo = \frac{\rho^{*}g^{*}R^{*2}}{\sigma^{*}},
\end{equation}\end{linenomath}
respectively.

Under scalings (\ref{eqn:Scalings}), the symmetry conditions (\ref{eqn:Dim_Symmetry_1}) become,
\refstepcounter{equation}
\begin{linenomath}
$$
 rhu = \frac{\partial h}{\partial r} = 0 \quad \mbox{at} \quad r = 0,
   \eqno{(\theequation{\mathit{a},\mathit{b}})}
 \label{eqn:NonDim_Symmetry_1}
$$
\end{linenomath}
while the contact line conditions (\ref{eqn:Dim_BCs_1}) are 
\refstepcounter{equation}
\begin{linenomath}
$$
 h = rhu = 0 \quad \mbox{at} \quad r = 1.
   \eqno{(\theequation{\mathit{a},\mathit{b}})}
 \label{eqn:NonDim_BCs_1}
$$
\end{linenomath}
The initial condition \eqref{eqn:Dim_ICs_1} becomes
\begin{linenomath}\begin{equation} 
h(r,0) = h_{0}(r) \quad \mbox{for} \quad 0<r<1.
 \label{eqn:NonDim_ICs_1}
\end{equation}\end{linenomath}
Finally, the dimensionless form of conservation of liquid volume conditions (\ref{eqn:Dim_Volume_1}) and (\ref{eqn:Dim_Volume_2}) is
\begin{linenomath}\begin{equation}
 1-t = 2\pi\int_{0}^{1}rh(r,t)\,\mbox{d}r.
 \label{eqn:NonDim_Volume}
\end{equation}\end{linenomath}

After scaling, the solute transport equation (\ref{eqn:Dim_Adv_Diff}) becomes
\begin{linenomath}\begin{equation}
 \frac{\partial}{\partial t}\left(h\phi\right) + \frac{1}{4r}\frac{\partial}{\partial r}\left[r\left(hu\phi - \frac{h}{\Pe}\frac{\partial\phi}{\partial r}\right)\right] = 0 
 \label{eqn:NonDim_Adv_Diff}
\end{equation}\end{linenomath}
for $0<r<$, $0<t<1$, where the solutal P\'{e}clet number is
\begin{linenomath}\begin{equation}
 \Pe = \frac{D^{*}(c_{s}^{*}-c_{\infty}^{*})}{\delta\rho^{*}D_{\phi}^{*}}.
\end{equation}\end{linenomath}
The symmetry (\ref{eqn:Dim_Symmetry_2}) and boundary conditions (\ref{eqn:Dim_BCs_2}) become
\begin{linenomath}\begin{equation}
 \frac{\partial \phi}{\partial r} = 0 \quad \mbox{at} \quad r = 0
 \label{eqn:NonDim_Symmetry_2}
\end{equation}\end{linenomath}
and
\begin{linenomath}\begin{equation}
 r\left(hu\phi - \frac{h}{\Pe}\frac{\partial \phi}{\partial r}\right) = 0 \quad \mbox{at} \quad r = 1,
 \label{eqn:NonDim_BCs_2}
\end{equation}\end{linenomath}
respectively. Finally, the initial condition (\ref{eqn:Dim_ICs_2}) becomes
\begin{linenomath}\begin{equation}
 \phi(r,0) = 1 \quad \mbox{for} \quad 0<r<1.
 \label{eqn:NonDim_ICs_2}
\end{equation}\end{linenomath}

\subsection{Integrated mass variable formulation}

The assumption that the solute is dilute decouples the flow and solute transport problems, so that we may solve for $h$ and $u$ from (\ref{eqn:NonDim_Lubrication_1})--(\ref{eqn:NonDim_Volume}) independently of the solute concentration, $\phi$. We shall discuss the resulting flow solution shortly in \textsection \ref{sec:Flow_solution}. 

First, however, we present a reformulation of the solute transport problem (\ref{eqn:NonDim_Adv_Diff})--(\ref{eqn:NonDim_ICs_2}), which will greatly aid us in our asymptotic and numerical investigations. In this, we follow \cite{Moore2021a,Moore2022} by introducing the \tit{integrated mass variable} 
\begin{linenomath}\begin{equation}
 \mathcal{M}(r,t) = \int_{0}^{r}sh(s,t)\phi(s,t) \,\mbox{d}s.
\end{equation}\end{linenomath}
By integrating the advection-diffusion equation (\ref{eqn:NonDim_Adv_Diff}) from $0$ to $r$ and applying the no-flux condition (\ref{eqn:NonDim_BCs_2}), we find that
\begin{linenomath}\begin{equation}
 \frac{\partial\mathcal{M}}{\partial t} + \left[\frac{u}{4} + \frac{1}{4\Pe}\left(\frac{1}{r} + \frac{1}{h}\frac{\partial h}{\partial r}\right)\right]\frac{\partial \mathcal{M}}{\partial r} - \frac{1}{4\Pe}\frac{\partial^{2}\mathcal{M}}{\partial r^{2}} = 0 \quad \mbox{for} \quad 0<r,t<1.
 \label{eqn:IMV_AdvDiff}
\end{equation}\end{linenomath}
This must be solved subject to the boundary conditions 
\refstepcounter{equation}
\begin{linenomath}
$$
 \mathcal{M}(0,t) = 0, \quad \mathcal{M}(1,t) = \frac{1}{2\pi}  \quad \mbox{for} \quad t>0, 
  \eqno{(\theequation{\mathit{a},\mathit{b}})}
 \label{eqn:IMV_BC}
$$
\end{linenomath}
where the latter condition dictates that mass is conserved along a radial ray, which replaces the no-flux condition (\ref{eqn:NonDim_BCs_2}). Finally, the initial condition (\ref{eqn:NonDim_ICs_2}) becomes
\begin{linenomath}\begin{equation}
 \mathcal{M}(r,0) = \int_{0}^{r}sh(s,0)\,\mbox{d}s \quad \mbox{for} \quad 0<r<1.
  \label{eqn:IMV_IC}
\end{equation}\end{linenomath}


Finally, we note that, once we have determined the integrated mass variable from (\ref{eqn:IMV_AdvDiff})--(\ref{eqn:IMV_IC}), the solute mass $m = \phi h$ can then be retrieved from
\begin{linenomath}\begin{equation}
 m = \frac{1}{r}\frac{\partial\mathcal{M}}{\partial r}.
 \label{eqn:SoluteMass}
\end{equation}\end{linenomath}

\section{Flow solution in the large-$\Ca$ limit} \label{sec:Flow}
\label{sec:Flow_solution}

We now suppose that surface tension dominates viscosity in the flow problem, that is $\Ca \gg 1$. Importantly, this means that the problems for the free surface profile and the flow velocity decouple, an assumption that is valid for a wide range of different liquids and evaporation models in practice \citep{Moore2021a,Moore2022}. Unlike these previous studies, however, we shall retain gravity in (\ref{eqn:NonDim_Lubrication_2}) to investigate what role it plays in the formation of the nascent coffee ring. 

To this end, we neglect the left-hand side of (\ref{eqn:NonDim_Lubrication_2}), so that upon integrating and applying the symmetry condition (\ref{eqn:NonDim_Symmetry_1}), the contact line condition (\ref{eqn:NonDim_BCs_1}a) and the conservation of liquid volume condition (\ref{eqn:NonDim_Volume}), we deduce that
\begin{linenomath}\begin{equation}
 h(r,t) = \frac{(1-t)}{\pi}\frac{I_{0}(\sqrt{\Bo})}{I_{2}(\sqrt{\Bo})}\left(1-\frac{I_{0}(\sqrt{\Bo}\,r)}{I_{0}(\sqrt{\Bo})}\right),
 \label{eqn:h}
\end{equation}\end{linenomath}
where $I_{\nu}(z)$ is the modified Bessel function of the first kind of order $\nu$.

With the free surface found, the velocity is determined immediately from (\ref{eqn:NonDim_Lubrication_1}) and the no-flux condition (\ref{eqn:NonDim_BCs_1}b) to be
\begin{linenomath}\begin{equation}
 u(r,t) = \frac{1}{rh}\left[\frac{2}{\pi}\sqrt{1-r^{2}} + \frac{4I_{0}(\sqrt{\Bo})}{\pi I_{2}(\sqrt{\Bo})}\left(\frac{r^{2}-1}{2} + \frac{1}{\sqrt{\Bo}I_{0}(\sqrt{\Bo})}(I_{1}(\sqrt{\Bo})-rI_{1}(\sqrt{\Bo}\,r))\right)\right].
 \label{eqn:u}
\end{equation}\end{linenomath}

Notably, as in the surface tension-dominated regime where $\Bo \rightarrow 0$, time is separable in both the free surface and fluid velocity profiles, and so merely acts to scale the functional form. In particular, this means that the streamlines and pathlines coincide, which we shall exploit when considering the regime in which solutal diffusion is negligible in \textsection \ref{sec:Secondary_Peak}.

\begin{figure}
\centering \scalebox{0.42}{\epsfig{file=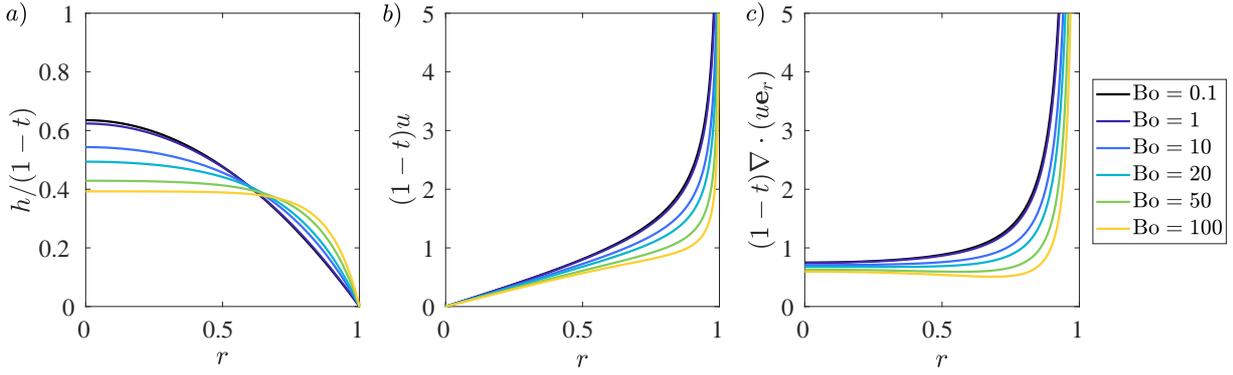}}
\caption{(a) The quasi-steady droplet free surface, (b) the fluid velocity, and (c) the divergence of the velocity displayed for $\Bo = 0.1$ (black), $1$ (dark purple), $10$ (blue), $20$ (cyan), $50$ (green) and $100$ (yellow). Notably, we see the transition from the spherical cap to the `pancake' droplet profile as the effect of gravity increases. The divergence of the fluid velocity also shows a transition from a monotonic to a non-monotonic profile as the Bond number increases.} 
\label{fig:FS_and_Velocity} 
\end{figure}

We display the scaled forms of the free surface and fluid velocity for various values of the Bond number in figure \ref{fig:FS_and_Velocity}a,b. For the droplet free surface profile, we see the expected transition from the spherical cap for $\Bo\rightarrow0$ \citep{Deegan2000} to the flat `pancake' droplet for $\Bo\rightarrow\infty$ \citep{rienstra1990shape}. For each Bond number, the velocity is singular at the contact line --- as expected for a diffusive evaporative flux (see, for example, \cite{Deegan2000}). We see that as the effect of gravity increases, the sharp increase in $u$ occurs closer to the contact line, corresponding to the progressively smaller region in which surface tension effects are important. 

Finally, since this will be important in our discussions of the secondary peaks seen in the solute mass profile in \textsection \ref{sec:Secondary_Peak}, we show the divergence of the fluid velocity in figure \ref{fig:FS_and_Velocity}c. For small Bond numbers, the divergence is monotonically increasing with $r$ and, as with the velocity, singular at the contact line. However, for moderate and large Bond numbers $\gtrsim 15$, we see a clear change of behaviour, with a region of non-monotonic behaviour in the droplet interior. This behaviour is accentuated as $\Bo\rightarrow\infty$.

For future reference, the asymptotic behaviours of the free surface and fluid velocity as $r\rightarrow1^{-}$ for $\Bo = O(1)$ are given by
\begin{linenomath}\begin{eqnarray}
h & = &  \theta_{c}(t;\Bo)(1-r) + O((1-r)^2), \label{eqn:h_local_Bo_unity}\\
 u & = &  \frac{2\chi}{\theta_{c}(t;\Bo)}(1-r)^{-1/2} + O\left((1-r)^{1/2}\right),
 \label{eqn:u_local_Bo_unity}
\end{eqnarray}\end{linenomath}
where
\begin{linenomath}\begin{equation}
\theta_{c}(t;\Bo) = -\lim_{r\rightarrow 1^-}\frac{\partial h}{\partial r} = (1-t)\psi(Bo), \quad \psi(\Bo) = \frac{\sqrt{\Bo}I_{1}(\sqrt{\Bo})}{\pi I_{2}(\sqrt{\Bo})}
\label{eqn:Contact_angle}
\end{equation}\end{linenomath}
is the leading order contact angle in the thin droplet limit and
\begin{linenomath}\begin{equation}
\chi = \frac{\sqrt{2}}{\pi}
\label{eqn:chi}
\end{equation}\end{linenomath}
is the dimensionless coefficient of the inverse square root singularity at the contact line in the evaporative flux (\ref{eqn:J}). Note that we have chosen this notation to highlight the similarities with the previous analysis of \cite{Moore2022}, who consider a surface tension-dominated droplet of arbitary contact set.


On the other hand, if we take $1-r = O(1)$ and consider the large-$\Bo$ limit of (\ref{eqn:h}), (\ref{eqn:u}), we find that
\begin{linenomath}\begin{eqnarray}
 h & = & h_{0}(t) + \Bo^{-1/2}h_{1}(t) + O(\Bo^{-1}), \\
 u & = & u_{0}(r,t) + \Bo^{-1/2}u_{1}(r,t) + O(\Bo^{-1}), 
\end{eqnarray}\end{linenomath}
as $\Bo\rightarrow\infty$, where
\refstepcounter{equation}
\begin{linenomath}$$
 h_{0}(t) = \frac{(1-t)}{\pi}, \quad  h_{1}(t) = \frac{2(1-t)}{\pi},
 \eqno{(\theequation{\mathit{a},\mathit{b}})}
 \label{eqn:h0}
$$\end{linenomath}
and
\refstepcounter{equation}
\begin{linenomath}$$
 u_{0}(r,t) = \frac{2\sqrt{1-r^{2}}}{r(1-t)}(1-\sqrt{1-r^{2}}), \quad u_{1}(r,t) = \frac{4}{r(1-t)}(1-\sqrt{1-r^{2}}).  
 \eqno{(\theequation{\mathit{a},\mathit{b}})}
 \label{eqn:u0} 
$$\end{linenomath}
Notably, in the droplet bulk, the droplet free surface $h$ is flat to all orders: the aforementioned characteristic of `pancake' droplets associated with large Bond numbers \citep{rienstra1990shape}. These expansions break down close to the contact line where surface tension effects become important. We find that for $1-r = \Bo^{-1/2}\bar{r}$, we have
\begin{linenomath}\begin{eqnarray}
 h & = & \bar{h}_{0}(\bar{r},t) + \Bo^{-1/2}\bar{h}_{1}(\bar{r},t) + O(\Bo^{-1}), \label{eqn:h_local_Bo_large} \\
 u & = & \Bo^{-1/4}\left[\bar{u}_{0}(\bar{r},t) + \Bo^{-1/4}\bar{u}_{1}(\bar{r},t) + \Bo^{-1/2}\bar{u}_{2}(\bar{r},t) + O(\Bo^{-3/4})\right]
 \label{eqn:u_local_Bo_large}
\end{eqnarray}\end{linenomath}
as $\Bo\rightarrow\infty$, where
\begin{linenomath}
\begin{alignat}{2}
 \bar{h}_{0}(\bar{r},t) & \; = && \; \frac{1}{\pi}(1-t)(1-\mbox{e}^{-\bar{r}}), \label{eqn:h0_inner}\\
 \bar{h}_{1}(\bar{r},t) & \; = && \; \frac{2(1-t)}{\pi}\left(1-\mbox{e}^{-\bar{r}}\right) - \frac{(1-t)\bar{r}}{2\pi}\mbox{e}^{-\bar{r}},
 \label{eqn:h1_inner}
\end{alignat}
\end{linenomath}
and
\begin{linenomath}
\begin{alignat}{2}
 \bar{u}_{0}(\bar{r},t) & \; = && \; \frac{2\sqrt{2\bar{r}}}{(1-t)(1-\mbox{e}^{-\bar{r}})}, \label{eqn:u0_inner}\\
 \bar{u}_{1}(\bar{r},t) & \; = && \; \frac{4}{(1-t)} - \frac{4\bar{r}}{(1-t)(1-\mbox{e}^{-\bar{r}})},\label{eqn:u1_inner}\\
 \bar{u}_{2}(\bar{r},t) & \; = && \; \frac{3\bar{r}^{3/2}}{\sqrt{2}(1-t)(1-\mbox{e}^{-\bar{r}})} - \frac{4\sqrt{2\bar{r}}}{(1-t)(1-\mbox{e}^{-\bar{r}})} + \frac{\sqrt{2}\bar{r}^{3/2}\mbox{e}^{-\bar{r}}}{(1-t)(1-\mbox{e}^{-\bar{r}})^{2}}.\label{eqn:u2_inner}
\end{alignat}
\end{linenomath}
We note here that as $\bar{r}\rightarrow0$, we retrieve the expect inverse square root singularity in the fluid velocity.

\section{Solute transport in the large-$\Pe$ limit}
\label{sec:Asymptotic_Analysis_alpha_O1}

Having fully determined the leading-order flow, we now seek to understand the transport of solute within the drop and to make predictions about the early-stages of coffee ring formation. We follow the analyses of \cite{Moore2021a,Moore2022} by considering the physically-relevant regime in which $\Pe \gg 1$. In this regime, in the bulk of the droplet, advection dominates solutal diffusion, with the latter only being relevant close to the contact line. 

Previous studies of this problem have concentrated on surface tension-dominated drops (i.e. $\Bo\rightarrow0$) and have shown how the competition between solutal advection and diffusion near the contact line leads to the early stages of coffee ring formation in drying droplets. In this analysis, we wish to investigate how this behaviour changes as we allow $\Bo$ to vary, which we pursue using a hybrid asymptotic-numerical approach.

There are naturally several different asymptotic regimes depending on the relative sizes of $\Bo$ and $\Pe$, but these broadly fall into two categories
\begin{enumerate}
    \item[i)] intermediate Bond number, $\Bo = O(1)$, where the asymptotic structure of the solute transport depends solely on the large P\'{e}clet number;
    \item[ii)] large Bond number, $\Bo\gg 1$, where the asymptotic structure of the solute transport now depends on the relative sizes of $\Bo$ and $\Pe$.  
\end{enumerate}

In the first regime where $\Bo = O(1)$, $\Pe\gg1$, the asymptotic structure of the flow is a natural extension of the surface tension-dominated case considered in \cite{Moore2021a}. In the droplet bulk where $1-r = O(1)$, solute advection dominates diffusion. However, close to the contact line, a balance between solute advection and diffusion occurs when
\begin{linenomath}\begin{equation}
 rhu\phi \sim \frac{rh}{\Pe}\frac{\partial\phi}{\partial r} \implies 1-r = O(\Pe^{-2}).
\end{equation}\end{linenomath}
We discuss the asymptotic solution for this regime in \textsection \ref{sec:Asymptotic_Analysis_Bo_moderate}.

In the second regime, there are several different possibilities depending on the relative sizes of the boundary layer where surface tension enters the flow profile and the solutal diffusion boundary layer. The richest distinguished asymptotic limit is that in which these boundary layers are comparable. As detailed in \textsection \ref{sec:Flow_solution}, for large Bond number the free surface is flat in the bulk of the droplet, with the effect of surface tension restricted to a boundary layer at the contact line of size $1 - r = O(\Bo^{-1/2})$, where $h = O(1)$ and $u = O(\Bo^{-1/4})$.
Turning to the solute transport equation (\ref{eqn:NonDim_Adv_Diff}), since $h$ is order unity and $u$ is square root bounded in this region, advection and diffusion are comparable when
\begin{linenomath}\begin{equation}
1 - r = O(\Pe^{-2/3}). \label{eqn:Local_balance}
\end{equation}\end{linenomath}
Hence, in the most general limit in which the size of the two boundary layers are comparable, we have 
\begin{linenomath}\begin{equation}
 \alpha = \Bo^{-1/2}\Pe^{2/3} = O(1). \label{eqn:alpha}
\end{equation}\end{linenomath}
The asymptotic analysis in this regime is somewhat more involved, so for brevity, we present the details in Appendix \ref{appendix:Large_Bo}. 

\subsection{Asymptotic solution when $\Bo = O(1)$}
\label{sec:Asymptotic_Analysis_Bo_moderate}

In this section, we present the asymptotic solution of the solute transport problem as $\Pe\rightarrow\infty$ when $\Bo = O(1)$. The analysis herein is a natural extension of \cite{Moore2021a}. For the purposes of this section, we shall use the concentration form of the advection-diffusion equation (\ref{eqn:NonDim_Adv_Diff})--(\ref{eqn:NonDim_ICs_2}) and, in particular, find the solution in terms of the solute mass $m = \phi h$, where $h$ is given by (\ref{eqn:h}).

\subsubsection{Outer region}

In the droplet bulk where $1-r = O(1)$, we seek a solution of the form $m = m_{0}(r,t) + O(\Pe^{-1})$ as $\Pe\rightarrow\infty$. Substituting into (\ref{eqn:NonDim_Adv_Diff}), (\ref{eqn:NonDim_ICs_2}), we find that
\begin{linenomath}
\begin{equation}
 \frac{\partial m_{0}}{\partial t} + \frac{1}{4r}\frac{\partial}{\partial r}(rm_{0}u) = 0 \quad \mbox{for} \quad 0<r<1, \, t>0
 \label{eqn:Bo_unity_outer_equation}
\end{equation}
\end{linenomath}
where $u$ is given by (\ref{eqn:u}), subject to $m(r,0) = h(r,0)$. This is the usual advection equation, with solution given by
\begin{linenomath}\begin{equation}
 m_{0}(r,t) = \frac{h(R,0)}{J(R,t)},
 \label{eqn:Bo_unity_outer_soln}
\end{equation}\end{linenomath}
where $R$ is the initial location of the point that is at $r$ at time $t$ and $J(R,t)$ is the Jacobian of the Eulerian-Lagrangian transformation, that satisfies Euler's identity,
\begin{linenomath}\begin{equation}
 \frac{\mbox{D}}{\mbox{D}t}(\log{J}) = \frac{1}{4r}\frac{\partial}{\partial r}(ru), \quad J(R,0) = 1,
\end{equation}\end{linenomath}
where $\mbox{D}/\mbox{D}t$ is the convective derivative. 

A straightforward asymptotic analysis of (\ref{eqn:Bo_unity_outer_equation}) reveals that \begin{linenomath}
    \begin{equation}
    u\frac{\partial m}{\partial r} \sim \frac{m}{r}\frac{\partial }{\partial r}(ru)
    \end{equation}
\end{linenomath}
as $r\rightarrow1^-$, so that $m_{0} = O(\sqrt{1-r})$ as $r\rightarrow1^-$, and hence the concentration $\phi_0$ is square root singular. This sharp local concentration increase necessitates the inclusion of a diffusive boundary layer.


\subsubsection{Inner region}
\label{sec:Moderate_inner}


Close to the contact line, we set
\begin{linenomath}
\begin{equation}
 r = 1 - \Pe^{-2}\hat{r}, \quad h = \Pe^{-2}\hat{h}, \quad u = \Pe\hat{u}, \quad m = \Pe^{2}\hat{m}, 
\end{equation}
\end{linenomath}
where the last scaling on the mass comes from global conservation of solute considerations \citep{Moore2021a}. We seek an asymptotic solution of the form $\hat{m} = \hat{m}_{0}(\hat{r},t) + O(\Pe^{-1})$ and find to leading order
\begin{linenomath}
\begin{equation}
 \frac{\partial}{\partial \hat{r}}\left[\left( \frac{2\chi}{\theta_{c}(t;\Bo)\sqrt{\hat{r}}} - \frac{1}{\hat{r}}\right)\hat{m}_{0} + \frac{\partial\hat{m}_{0}}{\partial\hat{r}}\right] = 0 \quad \mbox{in} \quad \hat{r}>0, \, t>0
 \label{eqn:Bo_Inner_1}
 \end{equation}
\end{linenomath}
such that
\begin{linenomath}
\begin{equation}
 \left( \frac{2\chi}{\theta_{c}(t;\Bo)\sqrt{\hat{r}}} - \frac{1}{\hat{r}}\right)\hat{m}_{0} + \frac{\partial\hat{m}_{0}}{\partial\hat{r}} = 0 \quad \mbox{for} \quad \hat{r} = 0.
 \label{eqn:Bo_Inner_2}
\end{equation}
\end{linenomath}

It is straightforward to show that the solution to (\ref{eqn:Bo_Inner_1})--(\ref{eqn:Bo_Inner_2}) is given by
\begin{linenomath}\begin{equation}
 \hat{m}_{0}(\hat{r},t) = C(t;\Bo)\hat{r}\mbox{exp}\left(-\frac{4\chi}{\theta_{c}(t;\Bo)}\sqrt{\hat{r}}\right),
 \label{eqn:Bo_unity_inner_soln}
\end{equation}\end{linenomath}
where, by pursuing a similar matching process to \citet{Moore2022}, we find that the coefficient $C(t;\Bo)$ is given by
\begin{linenomath}\begin{equation}
C(t) = \frac{64\chi^4}{3\theta_{c}(t;\Bo)^4}\mathcal{N}(t;\Bo),
\end{equation}\end{linenomath}
where $\mathcal{N}(t;\Bo)$ is the leading-order accumulated mass advected into the
contact line region up to time $t$, viz.
\begin{linenomath}
\begin{equation}
\mathcal{N}(t;\Bo) = \frac{1}{4}\int_{0}^{t} m_{0}(r,\tau)u(r,\tau)\,\mbox{d}\tau.
\label{eqn:MassFluxIn_Moderate_Bo}
\end{equation}
\end{linenomath}

\begin{figure}
\centering \scalebox{0.55}{\epsfig{file=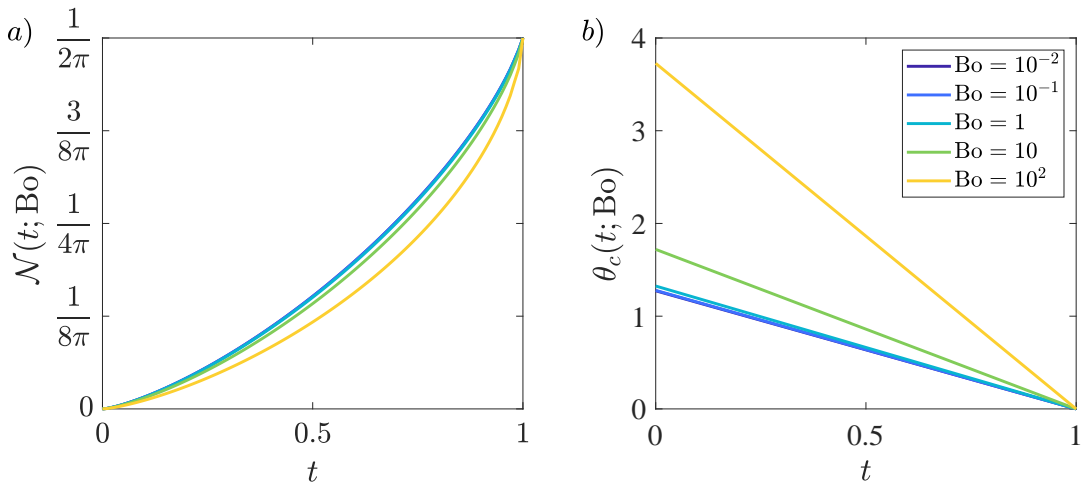}}
\caption{(a) The accumulated mass flux, $\mathcal{N}(t;\Bo)$ as defined by (\ref{eqn:MassFluxIn_Moderate_Bo}) and (b) the leading order local contact angle $\theta_{c}(t;\Bo)$ as defined by (\ref{eqn:Contact_angle}), for $\Bo = 10^{-2}$ (purple), $\Bo = 10^{-1}$ (purple) (dark blue), $\Bo = 1$ (light blue), $\Bo = 10$ (green) and $\Bo = 10^{2}$ (yellow).}
\label{fig:Mass_Flux_and_CA} 
\end{figure}

It is worth noting that this solution follows directly from the $\Bo = 0$ regime discussed in \cite{Moore2021a, Moore2022}, with the alterations due to gravity entering into the accumulated mass flux into the contact line and the leading order contact angle. In particular, we note that in the limit $\Bo\rightarrow0$, since $\psi = 4/\pi + O(\Bo)$, this yields the expected form found in the surface tension-dominated problem in \cite{Moore2022} (see \textsection 3.7.2 therein). We display the accumulated mass flux and the local contact angle for a wide range of Bond numbers in figure \ref{fig:Mass_Flux_and_CA}. We see that as the influence of gravity increases, the acccumulated mass flux into the contact line at a fixed percentage of the evaporation time is reduced from the surface tension-dominated regime. On the other hand, the local contact angle increases, commensurate  with the droplet profile transitioning from a spherical cap to a `pancake' droplet. We note that this combined behaviour leads to $C(t;\Bo)$ decreasing as $\Bo$ increases. We discuss how these findings impact coffee ring formation in more detail in \textsection \ref{sec:Primary_Peak_Moderate_Bo}.


\subsubsection{Composite solution}

We may use van Dyke's rule \citep{VanDyke1964} to formulate a leading-order composite solution for the solute mass that is valid throughout the drop by combining the leading-order-outer solution (\ref{eqn:Bo_unity_outer_soln}) and the leading-order-inner solution (\ref{eqn:Bo_unity_inner_soln}), finding
\begin{linenomath}\begin{equation}
 m_{\mathrm{comp}}(r,t) = m_{outer}(r,t) + \Pe^{2}m\left(\Pe^{2}(1-r),t\right)
 \label{eqn:Bo_unity_composite}.
\end{equation}\end{linenomath}


\begin{figure}
\begin{subfigure}
\centering \scalebox{0.5}{\epsfig{file=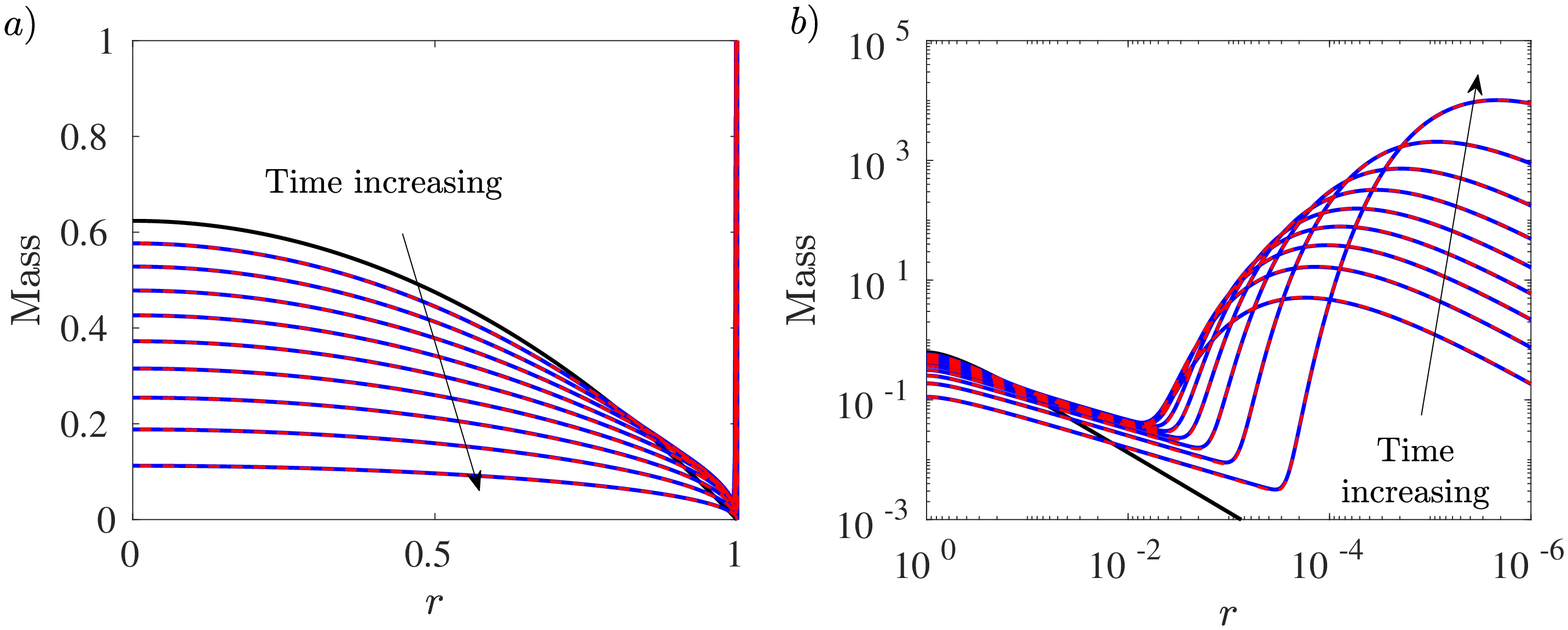}}
\captionsetup{labelformat=empty}
\end{subfigure}
\begin{subfigure}
\centering \scalebox{0.5}{\epsfig{file=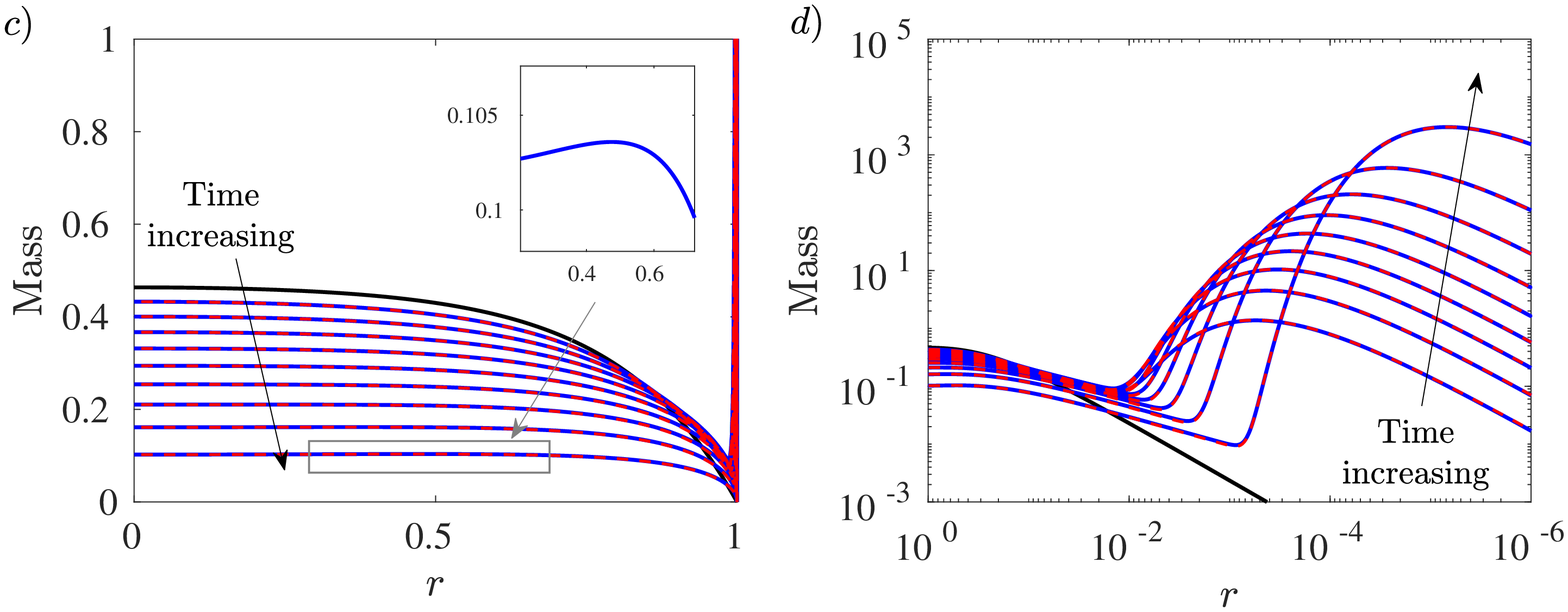}}
\end{subfigure}
\caption{Profiles of the solute mass when an axisymmetric droplet evaporates under a diffusive evaporative flux for (a,b) $\Pe = 10^{2}$ and $\Bo = 1$ (c,d) $\Pe = 10^2$ and $\Bo = 30$. In each figure, the bold, black curve represents the initial mass profile, which corresponds to the droplet free surface profile (\ref{eqn:h}). We also display plots at time intervals of 0.1 up to $t = 0.9$ in which solid, blue curves represent the results from the numerical solution of (\ref{eqn:IMV_AdvDiff})--(\ref{eqn:IMV_IC}) and the dashed, red curves show the leading-order composite mass profile, given by (\ref{eqn:Bo_unity_composite}). The right-hand figures display a close-up of the profiles near the contact line. In (c), the inset shows a close up of the mass profile in the droplet interior at $t = 0.9$ where we see a clear formation of a secondary peak.}
\label{fig:Composite_Comparisons_Moderate_Bo} 
\end{figure}

\begin{figure}
\begin{subfigure}
\centering \scalebox{0.5}{\epsfig{file=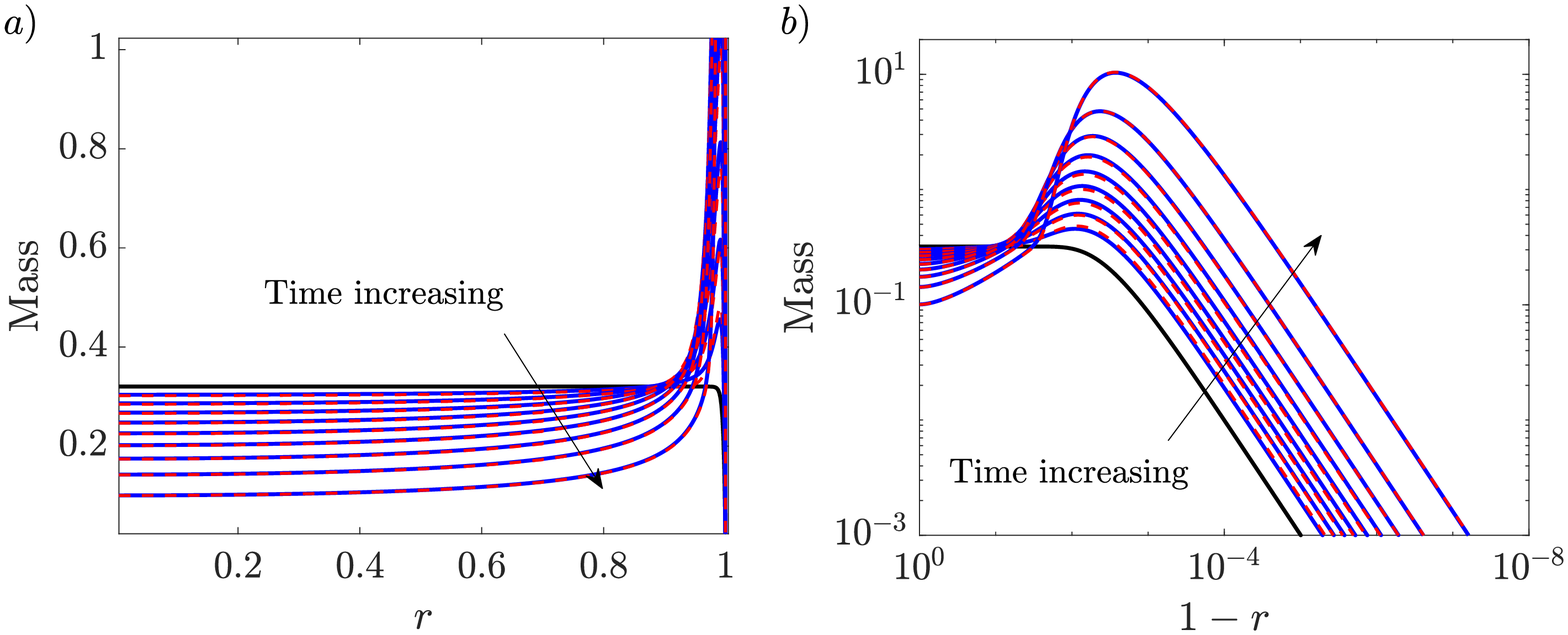}}
\captionsetup{labelformat=empty}
\end{subfigure}
\begin{subfigure}
\centering \scalebox{0.5}{\epsfig{file=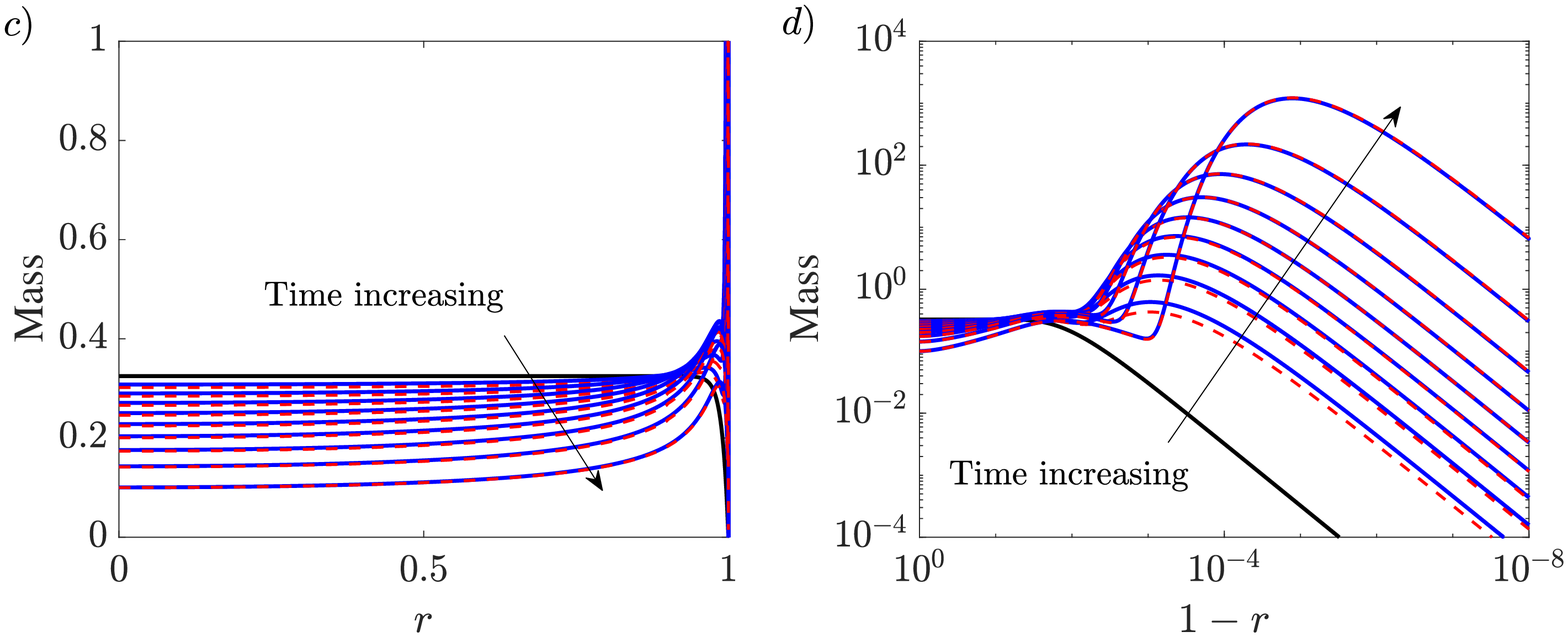}}
\end{subfigure}
\caption{Profiles of the solute mass when an axisymmetric droplet evaporates under a diffusive evaporative flux for (a,b) $\Pe = 10^{2}$ and $\Bo = 10^5$ ($\alpha \approx 0.07$) (c,d) $\Pe = 10^3$ and $\Bo = 10^4$ ($\alpha = 1$). In each figure, the bold, black curve represents the initial mass profile (\ref{eqn:IMV_IC}). We also display plots at time intervals of 0.1 up to $t = 0.9$ in which solid, blue curves represent the results from the numerical solution of (\ref{eqn:IMV_AdvDiff})--(\ref{eqn:IMV_IC}) and the dashed, red curves show the composite mass profiles, given by (\ref{eqn:IMV_composite}) for the integrate mass variable and (\ref{eqn:Mass_composite}) for the solute mass, respectively. Note that in (c,d), we can clearly see the development of the secondary peak behind the primary peak.}
\label{fig:Composite_Comparisons_Pe3_Bo_1e4} 
\end{figure}

\subsection{Comparisons between the numerical and asymptotic results}
\label{sec:Results_alpha_O1}

Our asymptotic predictions are compared to numerical simulations of the full advection-diffusion problem for the integrated mass variable given by (\ref{eqn:IMV_AdvDiff})--(\ref{eqn:IMV_IC}). The integrated mass variable is chosen over the solute mass $m$ or the concentration $\phi$ since it is better behaved close to the contact line. The numerical procedure requires careful consideration of the thin diffusive boundary layer and we follow a similar approach to that described for the surface tension-dominated problem by \cite{Moore2021a}. We give a summary of the methodologies in Appendix \ref{appendix:Numerics}. 


We begin by comparing the asymptotic predictions of the solute mass profiles to numerical solutions in the regime where $\Bo = O(1)$. In figure \ref{fig:Composite_Comparisons_Moderate_Bo}, we display asymptotic (dashed, red) and numerical (solid, blue) curves at 10\% intervals of the total drying time for $\Pe = 10^2$, $\Bo = 1$ (a,b) and $\Pe = 10^{2}$, $\Bo = 30$ (c,d). In each figure, we see excellent agreement between the simulations of the full system and the leading-order composite solution (\ref{eqn:Bo_unity_composite}). There is a clear formation of the expected coffee ring in the region near the contact line, where solutal diffusion and advection interact. We see that increasing the Bond number in this regime leads to a slight reduction of the size of the coffee ring. 

This behaviour is reminiscent of the $\Bo = 0$ regime considered previously by \cite{Moore2021a}. However, in the later stages of the $\Pe = 10^2$, $\Bo = 30$ example, we see evidence of a qualitative difference in behaviour, with the formation of another  peak in the mass profile in the droplet interior (see inset in figure \ref{fig:Composite_Comparisons_Moderate_Bo}(c)). Henceforth, we shall refer to the classical coffee ring as the \tit{primary peak} and this new feature as the \tit{secondary peak}. The presence of the secondary peak depends on the Bond number, as there is no secondary peak in any of the profiles when $\Bo = 1$, but it also depends on the drying time, as the peak only develops in the later stages of evaporation when $\Bo = 30$ (between $60-70\%$ of the drying time). Noticeably, the secondary peak is significantly smaller in magnitude than the primary peak. 


For larger Bond numbers, we compare the numerical results to the asymptotic predictions in Appendix \ref{appendix:Large_Bo}. In figure \ref{fig:Composite_Comparisons_Pe3_Bo_1e4}, we display results for $\Pe = 10^2, \Bo = 10^5$ ($\alpha \approx 0.07$) (a,b) and $\Pe = 10^3$ and $\Bo = 10^4$ ($\alpha = 1$) (c,d). In each case, we display the composite profile for the solute mass given by (\ref{eqn:Mass_composite}). In each figure, we see that after an initial transient the asymptotic predictions and numerical results are again in excellent agreement. Moreover, we see further evidence of the existence of a  secondary peak in the case $\Pe = 10^3, \Bo = 10^4$ regimes, where the peak appears much earlier and is noticeably larger than that in the previous example (cf. figure \ref{fig:Composite_Comparisons_Moderate_Bo}c, where $\Pe = 10^2$, $\Bo = 30$). However, we also note again the strong dependence of the secondary peak on $\Bo$ and, possibly, $\Pe$, as there is no evidence of such an interior peak when $\Pe = 10^2$, $\Bo = 10^5$.

These findings prompt us to investigate this new feature more closely, alongside a discussion of how the characteristics of the primary peak --- and hence the classical coffee ring --- depend on the Bond number. 

\section{Properties of the two peaks}
\label{sec:Peak_analysis}

Given the excellent comparisons displayed in the previous section, we seek to use our asymptotic results to investigate properties of the nascent coffee ring and, in particular, the new feature of these moderate-to-large Bond number regimes: the secondary peak.

\subsection{Primary peak}

We shall begin by discussing the effect of the Bond number on the primary peak. As in previous studies of the surface tension-dominated regime, the formation of the primary peak is driven by the competing diffusive and advective solute fluxes \citep{Moore2021a,Moore2022} and is always present in the large-$\Pe$ regime. Furthermore, since all of the features of interest are well within the solutal diffusion boundary layer, we will use the inner solution --- as discussed in \textsection \ref{sec:Moderate_inner} in the $\Bo = O(1)$ regime and \textsection \ref{sec:Inner} in the large-$\Bo$ regime --- to do this.

\subsubsection{$\Bo = O(1)$ regime}
\label{sec:Primary_Peak_Moderate_Bo}

When the Bond number is order unity, the analysis is a natural extension of that in \cite{Moore2021a,Moore2022}. The local solute profile is dominated by the leading-order inner solution (\ref{eqn:Bo_unity_inner_soln}). Introducing the time-dependent P\'{e}clet number
\begin{linenomath}
\begin{equation}
\Pe_{t} = \frac{\Pe}{1-t},
\end{equation}
\end{linenomath}
the nascent coffee ring profile may be seen to have the similarity form
\begin{linenomath}
\begin{equation}
 \frac{\hat{m}_{0}(R,t)}{\Pe_{t}^{2}\mathcal{N}(t;\Bo)} = \frac{2\chi}{3\psi(\Bo)}f\left(\sqrt{R},3,\frac{4\chi}{\psi(\Bo)}\right), \quad R = \Pe_{t}^{2}(1-r)
\label{eqn:Bo_Unity_similarity_profile}
\end{equation}
\end{linenomath}
where $\psi$ and $\chi$ retain their definitions from (\ref{eqn:Contact_angle}) and (\ref{eqn:chi}) as the initial local contact angle and the coefficient of the evaporative flux singularity, respectively, and $f(x,k,l) = l^kx^{k-1}\mbox{e}^{-lx}/\Gamma(k)$ is the probability density function of a gamma distribution. It is this functional form which describes the characteristic narrow, sharp peak of the coffee ring.

\begin{figure}
\centering \scalebox{0.525}{\epsfig{file=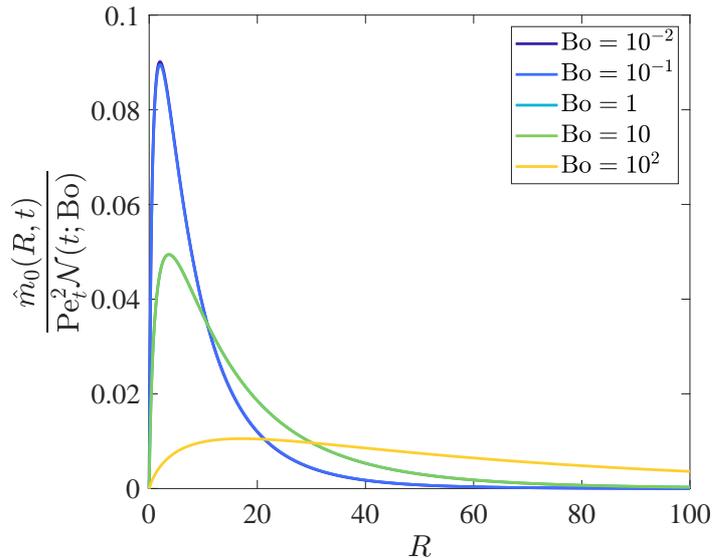}}
\caption{The similarity profile (\ref{eqn:Bo_Unity_similarity_profile}) of the leading-order-inner solute mass profile for $\Bo = 10^{-2}$ (purple), $\Bo = 10^{-1}$ (purple) (dark blue), $\Bo = 1$ (light blue), $\Bo = 10$ (green) and $\Bo = 10^{2}$ (yellow).}
\label{fig:Similarity_Profile} 
\end{figure}

Since the definition of $R$ only depends on the time-dependent P\'{e}clet number, we can clearly illustrate the effect of gravity by plotting the similarity profile (\ref{eqn:Bo_Unity_similarity_profile}) for a range of Bond numbers in figure \ref{fig:Similarity_Profile}. We see that, as the effect of gravity increases, the height of the primary peak decreases, and the peak moves further from the pinned contact line. Moreover, the shape of the primary peak tends towards a shallower, wider profile. 
Notably, this behaviour is driven purely by changes in $\psi(\Bo)$; as we saw in figure \ref{fig:Mass_Flux_and_CA}a, the accumulated mass flux into the contact line decreases with the Bond number, clearly this acts to accentuate this behaviour.

We can expand upon these results by finding the leading order asymptotic prediction of the primary peak height and location, which are given by
\begin{linenomath}
\begin{equation}
 r_{\mathrm{peak}, I}(t;\Bo) = 1 - \frac{\psi(\Bo)^2}{4\Pe_t^2\chi^2}, \quad m_{\mathrm{peak}, I}(t;\Bo) = \frac{16\Pe_{t}^2\mathcal{N}(t;\Bo)\chi^2}{3\mbox{e}^{2}\psi(\Bo)^2},
 \label{eqn:Primary_peak_properties}
\end{equation}
\end{linenomath}
respectively. Notably, while gravity only influences the location of the primary peak through the initial local contact angle, $\psi(\Bo)$, the height depends on gravity through both the contact angle and the accumulated mass flux, $\mathcal{N}(t;\Bo)$. In particular, referring back to figure \ref{fig:Mass_Flux_and_CA}, this means that gravity has a stronger effect on the peak height than its location. 

\begin{figure}
\begin{subfigure}
\centering \scalebox{0.4}{\epsfig{file=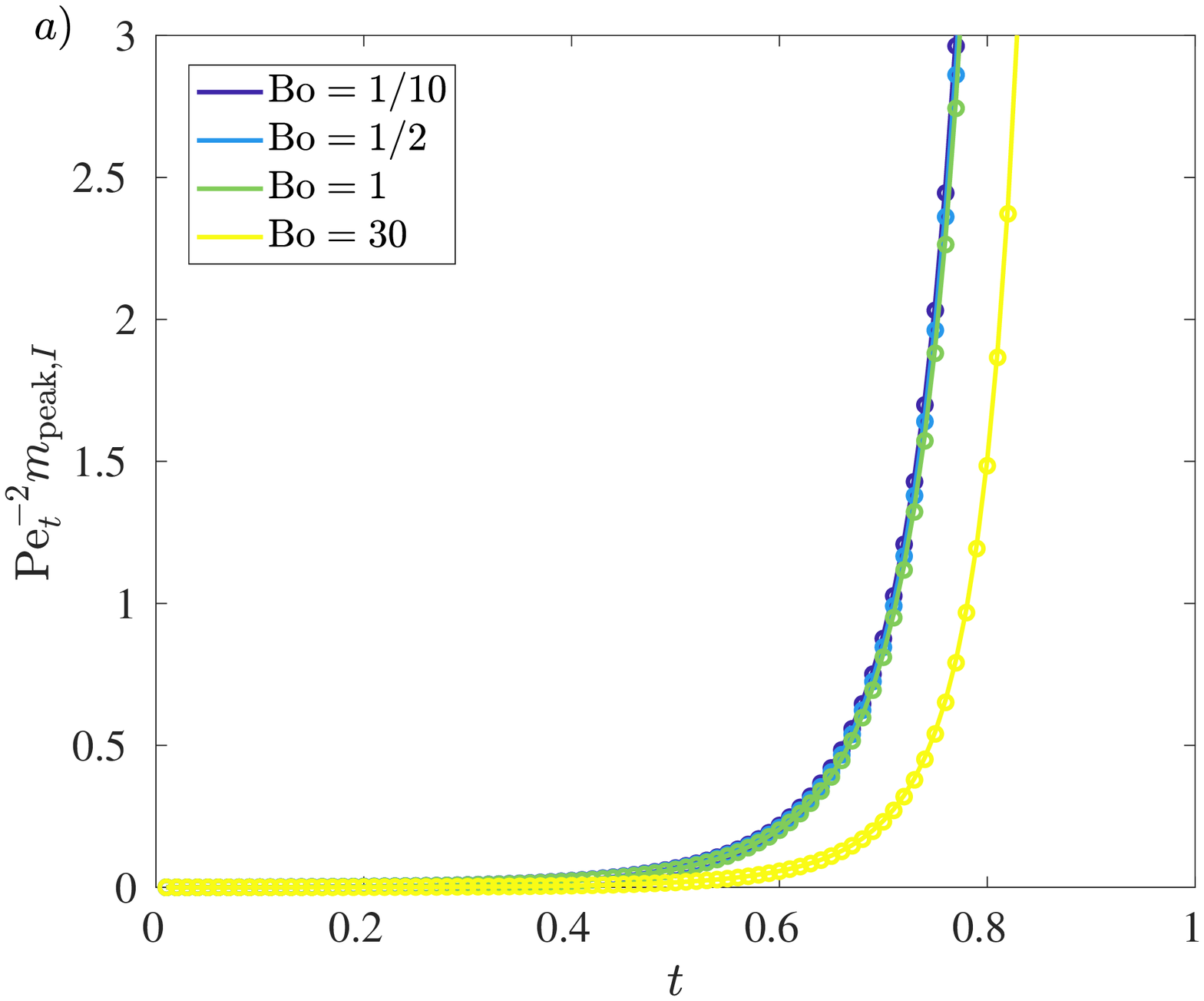}}
\captionsetup{labelformat=empty}
\end{subfigure}
\begin{subfigure}
\centering \scalebox{0.4}{\epsfig{file=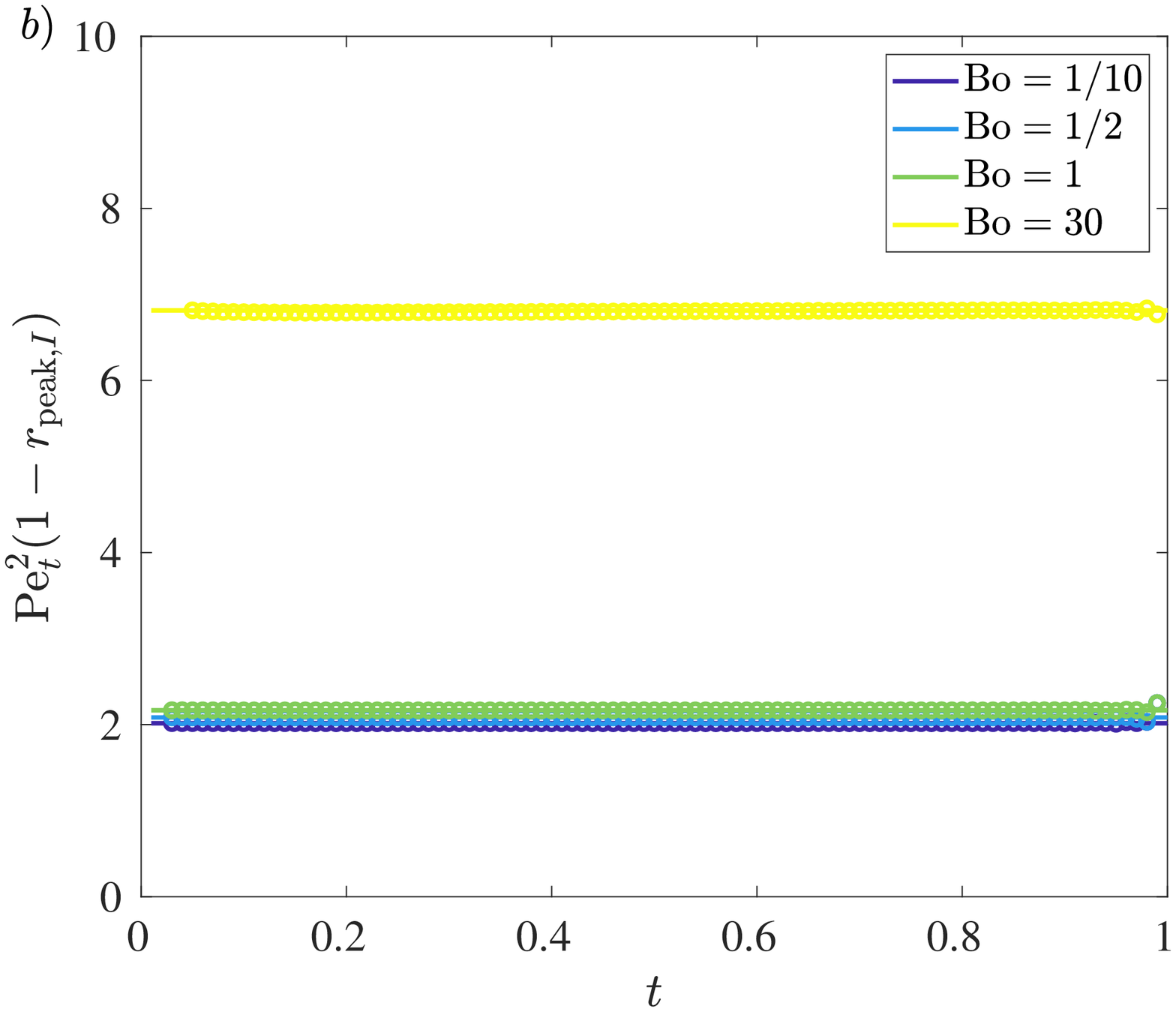}}
\end{subfigure}
\caption{Numerical (circles) and asymptotic predictions (solid lines) of ($a)$) the height of the primary peak, $m_{\mathrm{peak},I}(t)/\Pe_t^{2/3}$ and ($b)$) its location $\Pe_t^{2/3}(1-r_{\mathrm{peak},I}(t))$ in the $\Bo = O(1)$ regime as given by (\ref{eqn:Primary_peak_properties}). For each curve, $\Pe = 10^2$, while the Bond number varies according to $\Bo = 1/10$ (dark purple), $\Bo = 1/2$ (blue), $\Bo = 1$ (green) and $\Bo = 30$ (yellow).}
\label{fig:Primary_Ring_Properties_Bo_unity} 
\end{figure}

We illustrate the veracity of these asymptotic predictions by comparing them to the  corresponding numerical results for $\Pe = 10^2$ and a range of Bond numbers in figure \ref{fig:Primary_Ring_Properties_Bo_unity}. As anticipated from the comparisons of the solute mass profiles, we see excellent agreement between the asymptotic predictions and the numerical results. In particular, in figure \ref{fig:Primary_Ring_Properties_Bo_unity}a, we note that as the influence of gravity increases (i.e. $\Bo$ increases), the coffee ring effect is inhibited: although a peak clearly still forms, it is lower for large Bond number at a similar stage of the drying process. This effect varies nonlinearly with time (cf. figure \ref{fig:Mass_Flux_and_CA}a). For example, considering the cases $\Bo = 1/2$ and $\Bo = 30$, after $50\%$ of the drying time, the peak height is reduced by a factor of $\approx 3.97$, while at $60\%$ of the drying time, the reduction is a factor of $\approx 3.85$ and at $90\%$ of the drying time, it is $\approx 3.63$.

Similarly, in figure \ref{fig:Primary_Ring_Properties_Bo_unity}b, we see that as the Bond number increases, the location of primary peak moves further from the contact line and that this significantly increases as the Bond number gets larger. For $\Bo = 1/10, 1/2, 1$ the location is almost indistinguishable from the zero-Bond number solution --- where $\Pe_t^{2}(1-r) = 2$ (\cite{Moore2022}) --- but for $\Bo = 30$, this has increased to $\approx 6.79$.

It is worth noting that in all this analysis, the P\'{e}clet number simply acts to scale the above findings. For a larger P\'{e}clet number, the height of the primary peak increases, while it is located closer to the contact line. This is precisely what is seen for the $\Bo = 0$ regime \citep{Moore2021a}.  

\subsubsection{Large-$\Bo$ regime}

\begin{figure}
\begin{subfigure}
\centering \scalebox{0.4}{\epsfig{file=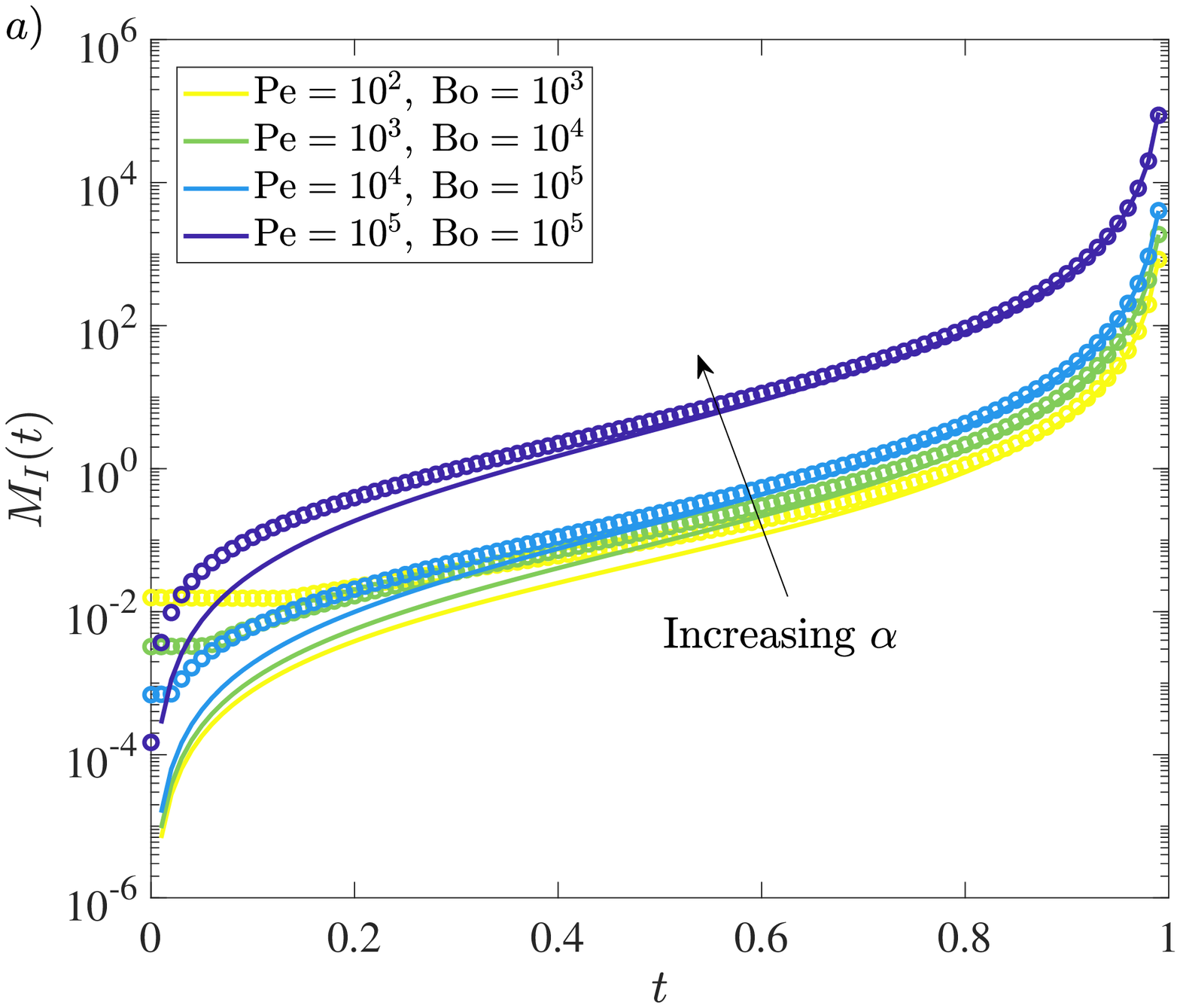}}
\captionsetup{labelformat=empty}
\end{subfigure}
\begin{subfigure}
\centering \scalebox{0.4}{\epsfig{file=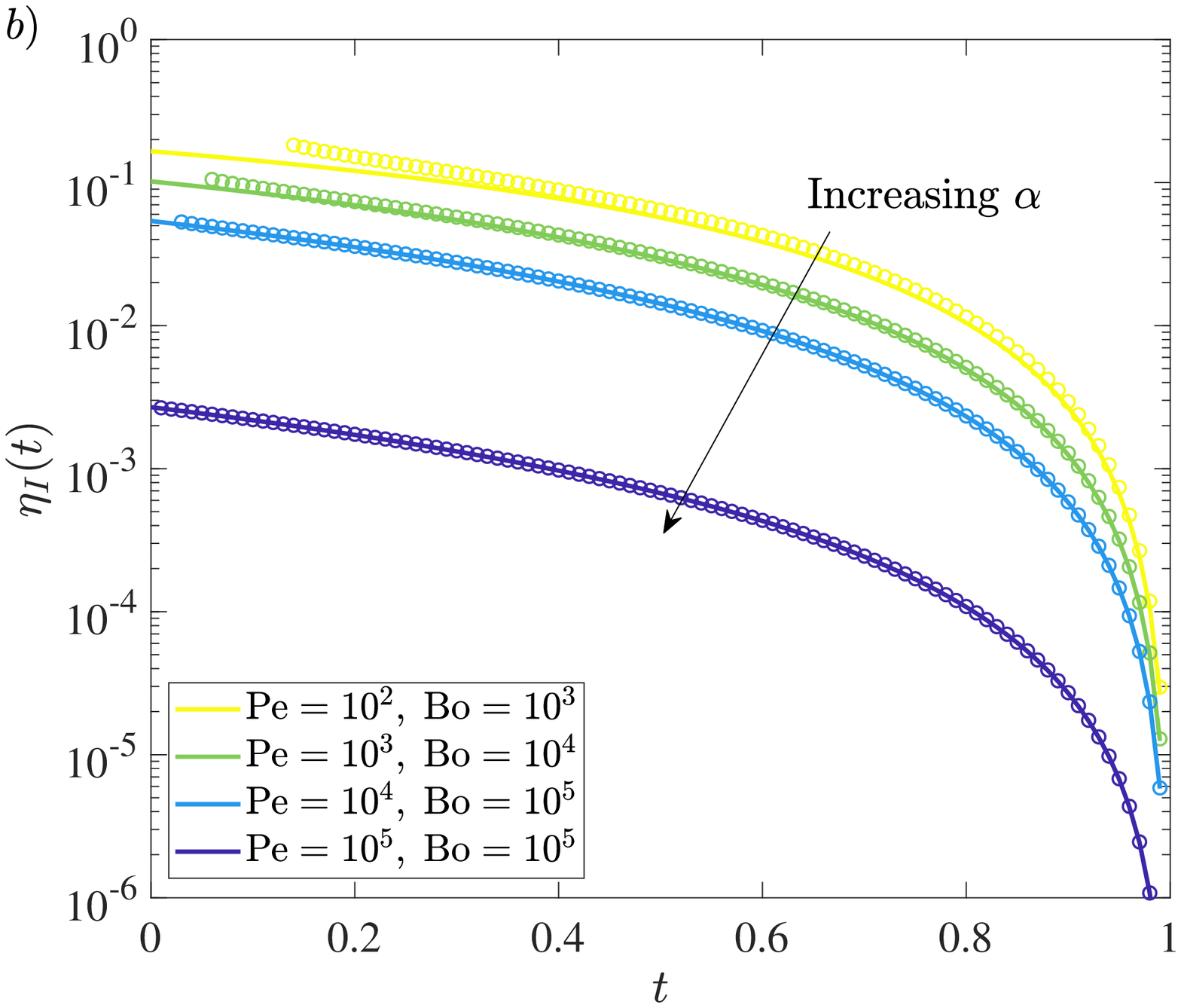}}
\end{subfigure}
\caption{Numerical (circles) and asymptotic predictions (solid lines) of ($a)$) the height of the primary peak, $M_{I}(t) = m_{\mathrm{peak},I}(t)/\Pe^{2/3}$ and ($b)$) its location $\eta_{I}(t) = \Pe^{2/3}(1-r_{\mathrm{peak},I}(t))$ as given by (\ref{eqn:Primary_Peak_Location})--(\ref{eqn:Primary_Peak_Height}). Results are presented for $\Pe = 10^{2}, \Bo = 10^{3}$ ($\alpha \approx 0.68$, yellow), $\Pe = 10^{3}, \Bo = 10^{4}$ ($\alpha = 1$, green), $\Pe = 10^{4}, \Bo = 10^{5}$ ($\alpha \approx 1.47$, blue) and $\Pe = 10^{5}, \Bo = 10^{5}$ ($\alpha \approx 6.81$, dark purple).}
\label{fig:Primary_Ring_Properties} 
\end{figure}

In the large-$\Bo$ regime, given the size of the primary peak, we anticipate that the leading-order-inner solution $\tM_{0}(\tr,t)$ as given by (\ref{eqn:IMV_inner_LO}) should reasonably capture the features of the primary peak. However, unlike its moderate-$\Bo$ counterpart, there is no simple similarity form for the solution in this regime, so that we proceed more carefully. 


We denote the height and location of the primary peak by 
\refstepcounter{equation}
\begin{linenomath}
$$
 m_{\mathrm{peak},I}(t) = \Pe^{2/3}M_{I}(t), \quad r_{\mathrm{peak},I}(t) = 1-\Pe^{-2/3}\eta_{I}(t),
  \eqno{(\theequation{\mathit{a},\mathit{b}})}
$$
\end{linenomath}
respectively. By \eqref{eqn:SoluteMass}, the location of the maximum $\eta_{I}(t)$ satisfies
\begin{linenomath}\begin{equation}
 0 = \frac{\partial^{2}\tM_{0}}{\partial\tr^{2}}(\eta_{I}(t),t).
\end{equation}\end{linenomath}
Utilizing (\ref{eqn:IMV_inner_LO_problem}), we find that
\begin{linenomath}\begin{equation}
 \frac{\partial^{2}\tM_{0}}{\partial\tr^{2}}(\eta_{I}(t),t) = -\left.\left(\tu_{0}-\frac{1}{\th_{0}}\frac{\partial\th_{0}}{\partial\tr}\right)\frac{\partial\tM_{0}}{\partial\tr}\right|_{(\eta_{I}(t),t)} = 0.
\end{equation}\end{linenomath}
Since $\partial\tM_{0}/\partial\tr > 0$ for $\tr>0$, we conclude 
\begin{linenomath}\begin{equation}
\tu_{0}(\eta_{I}(t),t) - \frac{1}{\th_{0}(\eta_{I}(t),t)}\frac{\partial\th_{0}}{\partial\tr}(\eta_{I}(t),t) = 0 
\end{equation}\end{linenomath}
so that
\begin{linenomath}\begin{equation}
\eta_{I}(t) = \frac{\alpha}{2}W_{0}\left(\frac{(1-t)^{2}}{4\alpha^{3}}\right),
\label{eqn:Primary_Peak_Location}
\end{equation}\end{linenomath}
where $W_{0}(x)$ is the Lambert-W function (i.e. the solution to $w\mbox{e}^{w} = x$).

With $\eta_{I}(t)$ in hand, the corresponding height of the ring at the peak is then given by
\begin{linenomath}\begin{equation}
 M_{I}(t) = \left(\frac{-B_{0}(t)}{I(\eta_{I}(t),t)}\right) = \left[\frac{\mathcal{N}(t)}{I(\eta_{I}(t),t)}\left(\int_{0}^{\infty}\frac{1}{I(s,t)}\,\mbox{d}s\right)^{-1}\right],
 \label{eqn:Primary_Peak_Height}
\end{equation}\end{linenomath}
where $I(r,t)$ is given by (\ref{eqn:Integrating_Factor}) and $\mathcal{N}(t)$ is the leading-order accumulated mass flux into the boundary layer (\ref{eqn:MassFluxIn}). Note that, in this regime, $\mathcal{N}(t)$ is independent of $\alpha$ and, hence, the Bond number, but the function $I(r,t)$ does change with $\alpha$.

In figure \ref{fig:Primary_Ring_Properties}, we plot the asymptotic predictions of the location and height of the primary deposit peak against the simulation results for a range of different P\'{e}clet and Bond numbers (and, correspondingly, $\alpha$). There are several discernible features. After an initial transient, the location of the peak is captured extremely well by the asymptotic prediction (\ref{eqn:Primary_Peak_Location}) for each case presented. This initial transient is primarily due to the lack of a distinct peak at early stages of the drying process; a period of time is necessary for sufficient solute to be advected to the contact line. This process takes longer for smaller P\'{e}clet numbers, i.e. when diffusion is relatively stronger. The height of the primary peak is captured quite well by the asymptotic prediction (\ref{eqn:Primary_Peak_Height}), particularly for larger P\'{e}clet numbers and as time increases. It is worth noting that the error in the approximation of the height is $O(\Pe^{1/3})$, so for an improved estimation of the primary peak height, it would be necessary to consider the first two inner solutions $\tM_{0}(\tr,t)$ and $\tM_{1}(\tr,t)$. While this is possible, the results do not have a simple analytic form, so are not practical to work with. We also note that, as the droplet evaporates, the primary peak both increases in size and moves closer to the contact line, i.e. $M_{I}(t)$ increases and $\eta_{I}(t)$ decreases as $t$ increases.

\subsection{Secondary peak}
\label{sec:Secondary_Peak}

As evidenced by the solute mass profiles, the behaviour of the secondary peak --- and indeed, even its presence --- is more complex than that of the primary peak, which always forms in the large-$\Pe$ regime. We have seen, for example in figure \ref{fig:Composite_Comparisons_Moderate_Bo} in the $\Bo = O(1)$ regime, that the presence of the peak varies with both $\Bo$ and drying time, while when $\Bo \gg 1$, we have also seen variation with $\Pe$ (and hence $\alpha$), see for example figure \ref{fig:Composite_Comparisons_Pe3_Bo_1e4}. This gives a clear indication that we need to treat this feature more carefully.

To begin, we will consider whether or not the secondary peak is present. We shall first fix the P\'{e}clet number and use the numerical results to produce a regime diagram in $(\Bo,t)$-parameter space indicating whether one or two peaks are present in the solute mass profile. We note here that these are the only options that we have been able to find --- we have found no instances of more than two peaks appearing. 

\begin{figure}
\begin{subfigure}
\centering \scalebox{0.55}{\epsfig{file=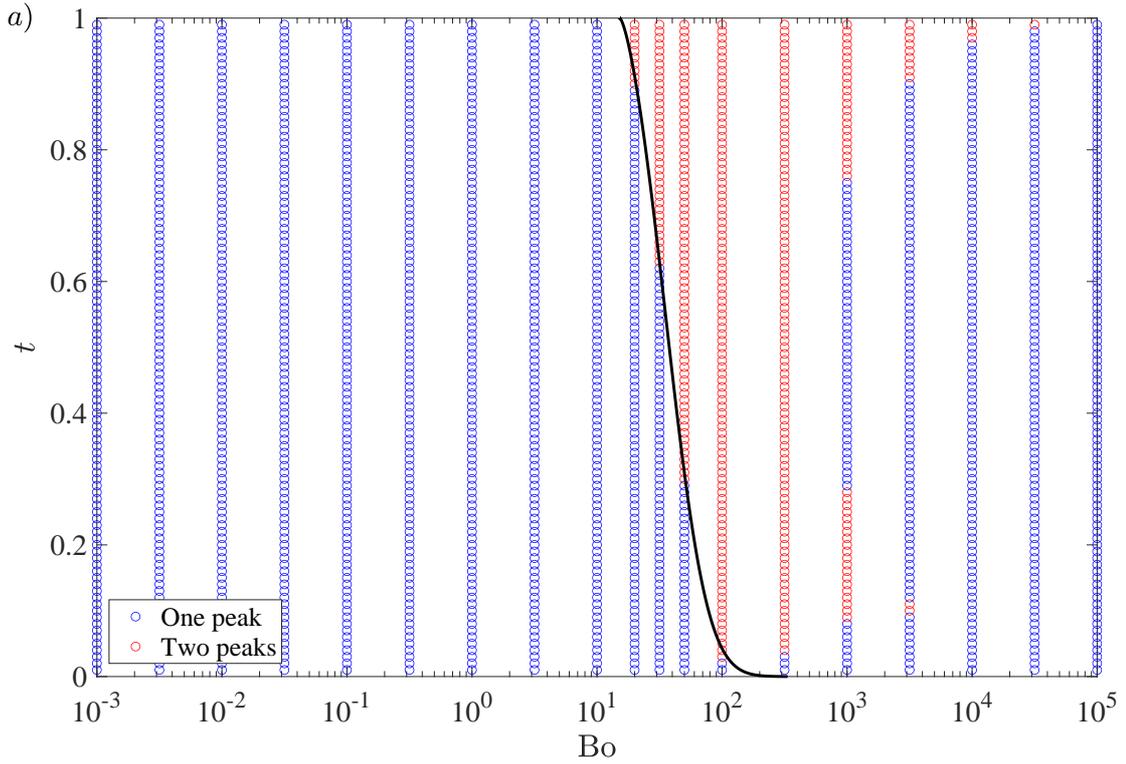}}
\captionsetup{labelformat=empty}
\end{subfigure}
\begin{subfigure}
\centering \scalebox{0.55}{\epsfig{file=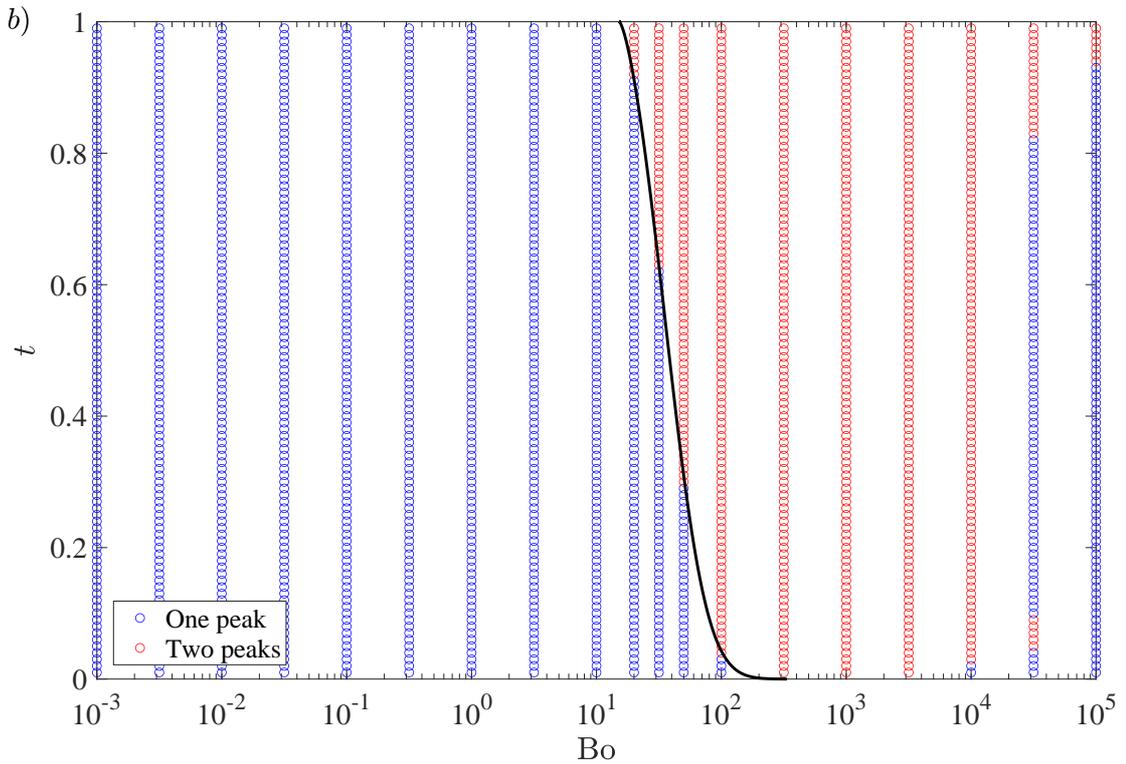}}
\end{subfigure}
\caption{$(\Bo,t)$-regime diagram illustrating the presence of either one (blue circles) or two (red circles) peaks in the solute mass profile for (a) $\Pe = 10^2$ and (b) $\Pe = 10^3$. The data is extracted from the numerical simulations and demonstrates a clear band of Bond numbers for which two peaks may exist in the profile. In each figure, the black curve denotes the asymptotic prediction of when the centre of the droplet changes from a maximum to a minimum as given by (\ref{eqn:Critical_tc}).}
\label{fig:Secondary_peak_Regime_diagram}
\end{figure}

We show the results for $\Pe = 10^2$ in figure \ref{fig:Secondary_peak_Regime_diagram}a. In the figure, solute profiles with one peak --- i.e. only the classical coffee ring --- are denoted by blue circles, while solute profiles exhibiting two peaks are denoted by red circles. We see a strong nonlinear dependence on both Bond number and dryout time. In particular, there is a band of Bond numbers between around $\Bo \approx 10$ and $\Bo \approx 30000$ that may lead to secondary peak formation, although the existence of a peak also depends strongly on $t$ for a fixed Bond number. We note that for $\Bo \lesssim 10$, there is only one peak for any $t$, in agreement with the classical $\Bo = 0$ regime. Moreover, for very large Bond number $\Bo \gtrsim 30000$, again we see that there is only one peak. 

We illustrate the effect of the P\'{e}clet number by plotting the equivalent regime diagram for $\Pe = 10^3$ in figure \ref{fig:Secondary_peak_Regime_diagram}b. Remarkably, the onset of the secondary peak appears to be unaffected by the increase of the P\'{e}clet number, although the band of Bond numbers for which we see two peaks is significantly widened into larger $\Bo$. Notably, however, the shape of the curve delineating between two peaks / one peak for large Bond number appears to be independent of $\Pe$, only its location has shifted.

\subsubsection{Onset of the secondary peak}

In this section, we seek to investigate some of the phenomena around the onset of the secondary peak in more detail. We saw that for a fixed P\'{e}clet number, there was a distinct switch from one to two peaks for Bond number $\Bo \approx 10$ and that this switch appears to be independent of $\Pe$. This suggests that secondary peak formation is not a result of the interplay between solutal advection and diffusion that drives the classical coffee ring. 


\begin{figure}
\centering \scalebox{0.5}{\epsfig{file=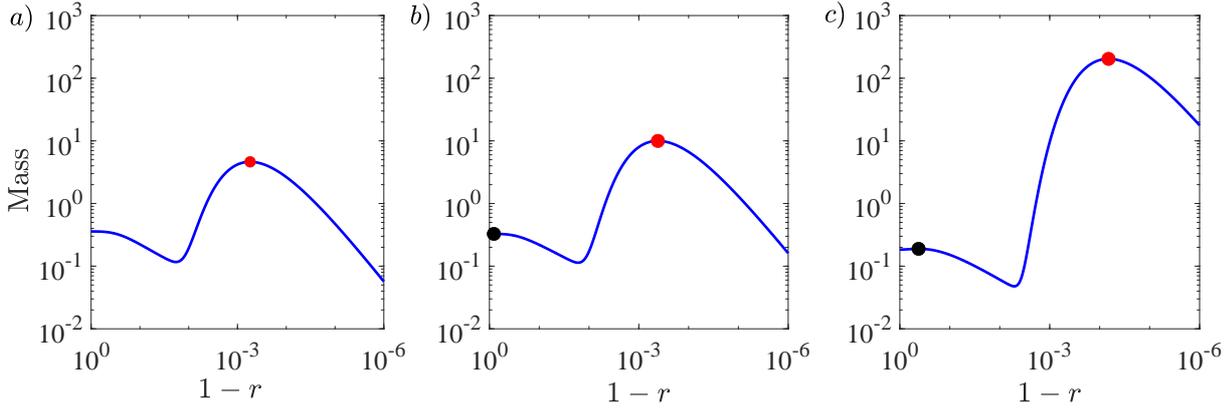}}
\caption{Solute profiles for an evaporating droplet with $\Pe = 10^{2}$ and $\Bo = 20$. The deposit profile is displayed on a doubly-logarithmic plot at $25\%$ ($a)$), $35\%$ ($b)$) and $75\%$ ($c)$) of the drying time in order to catch the emergence of the secondary peak. In each of $a)-c)$, the primary peak is indicated by a red circle, while the secondary peak is indicated by a black circle (when it exists).}
\label{fig:Secondary_Peak_Counterexample} 
\end{figure}

In order to investigate the reasons behind the presence of or lack of a secondary peak,  in figure \ref{fig:Secondary_Peak_Counterexample}, we plot numerical results for the solute profiles in a droplet with $\Pe = 10^{2}, \Bo = 20$ at $25\%, 35\%$ and $75\%$ of the drying time. In the figure, the primary and secondary peaks are indicated by the red and black circles, respectively. We clearly see in figure \ref{fig:Secondary_Peak_Counterexample}a that at $25\%$ of the drying time there is only one peak, but by 35\% of the drying time, the secondary peak has emerged close to the droplet centre. As the droplet evaporates further to $75\%$ of the drying time the secondary peak has moved further towards the droplet contact line.

This particular example gives us a strong indication that the secondary peak initially arises from the \tit{centre} of the drop and, in particular, appears to be linked with a transition from the centre being a \tit{maximum} in solute mass profile --- as it is for the classical coffee ring of \cite{Deegan1997,Deegan2000} --- to a \tit{minimum}.

To investigate this postulate, we consider the behvaiour close to the droplet centre. To simplify things, since the initial emergence of the secondary peak appears to be independent of the P\'{e}clet number, we neglect solutal diffusion completely, taking $\Pe = \infty$, so that the solute mass $m$ satisfies the first-order semi-linear equation
\begin{linenomath}\begin{equation}
 \frac{\partial m}{\partial t} + \frac{1}{4r}\frac{\partial}{\partial r}\left(rmu\right) = 0, \quad m(r,0) = h(r,0),
 \label{eqn:Secondary_Peak_Pe_inf}
\end{equation}\end{linenomath}
where, since the emergence appears to be rooted in the region where $\Bo \approx 10$, we consider the moderate Bond number regime and retain the full expressions for the droplet free surface $h$ and fluid velocity $u$ given by (\ref{eqn:h})--(\ref{eqn:u}). 

We seek an asymptotic solution of (\ref{eqn:Secondary_Peak_Pe_inf}) as $r\rightarrow0$. First, we note that for small arguments, the free surface and velocity have the following asymptotic expansions:
\begin{linenomath}
\begin{align}
 h(r,t) & \sim (1-t)\left[\mathcal{H}_{0}(\Bo) + \mathcal{H}_{1}(\Bo)r^{2} + o(r^2)\right], \\
 u(r,t) & \sim \frac{1}{(1-t)}\left[\mathcal{U}_{0}(\Bo)r + \mathcal{U}_{1}(\Bo)r^{3} + o(r^{3})\right]
\end{align}
\end{linenomath}
as $r\rightarrow0$, where
\begin{linenomath}
\begin{align}
 \mathcal{H}_{0}(\Bo) & = \frac{(I_{0}(\sqrt{Bo})-1)}{\pi I_{2}(\sqrt{\Bo})}, \\
 \mathcal{H}_{1}(\Bo) & = -\frac{\Bo}{4\pi I_{2}(\sqrt{\Bo})}, \\
 \mathcal{U}_{0}(\Bo) & = \frac{2\sqrt{\Bo} - \sqrt{\Bo}I_{0}(\sqrt{\Bo}) - 2I_{1}(\sqrt{\Bo})}{\sqrt{\Bo}(1-I_{0}(\sqrt{\Bo}))}, \\
 \mathcal{U}_{1}(\Bo) & = -\frac{(\Bo^{3/2}-\sqrt{\Bo}I_{0}(\sqrt{\Bo}) + \sqrt{\Bo}I_{0}(\sqrt{\Bo})^2 + 2I_{1}(\sqrt{\Bo})-2\Bo I_{1}(\sqrt{\Bo}) - 2I_{0}(\sqrt{\Bo})I_{1}(\sqrt{\Bo})}{4\sqrt{\Bo}(1-I_{0}(\sqrt{\Bo}))^{2}}.
\end{align}
\end{linenomath}

Now, by the symmetry of the problem, the droplet centre must be a critical point, so we seek a solution of the form $m = m_{0}(t) + m_{1}(t)r^{2}+o(r^{2})$ as $r\rightarrow0$. Upon substituting this ansatz and the above forms for $h$ and $u$ into (\ref{eqn:Secondary_Peak_Pe_inf}), straightforward calculation yields
\begin{linenomath}
\begin{align}
 m_{0}(t) & = \mathcal{H}_{0}(1-t)^{\mathcal{U}_{0}/2}, \\
 m_{1}(t) & = \left(\frac{2\mathcal{U}_{1}\mathcal{H}_{0}}{\mathcal{U}_{0}} + \mathcal{H}_{1}\right)(1-t)^{\mathcal{U}_{0}} - \frac{2\mathcal{U}_{1}\mathcal{H}_{0}}{\mathcal{U}_{0}}(1-t)^{\mathcal{U}_{0}/2}.
\end{align}
\end{linenomath}

Hence, given that initially the droplet has a maximum at its centre for any $\Bo$, we deduce that the maximum becomes a minimum at the critical time $t_{c}$ such that
\begin{linenomath}\begin{equation}
 m_{1}(t_{c}) = 0.
 \label{eqn:Secondary_peak_critical_condition}
\end{equation}\end{linenomath}
Since $2\mathcal{U}_{1}\mathcal{H}_{0}/\mathcal{U}_{0} + \mathcal{H}_{1}<0$, $\mathcal{H}_{0}>0$, $\mathcal{U}_{0}>0$ for all $\Bo$, (\ref{eqn:Secondary_peak_critical_condition}) only has solutions for $\Bo > \Bo_{c}$ where
\begin{linenomath}\begin{equation}
\mathcal{U}_{1}(\Bo_{c}) = 0 \quad \implies \quad \Bo_{c}\approx 15.21.
\label{eqn:Critical_Bo}
\end{equation}\end{linenomath}
When $\Bo>\Bo_c$, we may solve (\ref{eqn:Secondary_peak_critical_condition}) explicitly to find
\begin{linenomath}\begin{equation}
t_{c}(\Bo) = 1 - \left(\frac{2\mathcal{U}_{1}(\Bo)\mathcal{H}_{0}(\Bo)}{2\mathcal{U}_{1}(\Bo)\mathcal{H}_{0}(\Bo)+\mathcal{H}_{1}(\Bo)\mathcal{U}_{0}(\Bo)}\right)^{2/\mathcal{U}_{0}(\footnotesize{\Bo})}.
\label{eqn:Critical_tc}
\end{equation}\end{linenomath}

This critical curve in figure \ref{fig:Secondary_peak_Regime_diagram} is displayed as the solid black curve and we see that there is excellent agreement between this prediction and the transition from one to two peaks. But, what is causing the transition? Since the phenomenon is independent of the P\'{e}clet number, it is purely a result of the droplet geometry and the evaporation-driven flow. In particular, we note that the critical Bond number $\Bo_{c}$ given by (\ref{eqn:Critical_Bo}) is linked to the change in sign of $\mathcal{U}_{1}$, which is equivalent to requiring that $(1-t)\nabla\cdot(u\mbf{e}_{r})$ is \tit{decreasing} near $r = 0$. This correlates with the profiles of the divergence of $\mbf{u}$ displayed in figure \ref{fig:FS_and_Velocity}c, where we see this change in sign clearly as the Bond number increases. 

Notably, considering the curve displayed in figure \ref{fig:Secondary_peak_Regime_diagram}, we see that for $\Bo$ close to $\Bo_{c}$, the secondary peak only emerges very late in the dryout process, but as the Bond number increases, it appears almost instantaneously. Hence, from this analysis alone, we might expect there to always be two peaks for $\Bo>\Bo_{c}$, but clearly this is not the case. We now investigate why in more detail.


\subsubsection{Loss of the secondary peak}

Given its clear variation with each of $t$, $\Bo$ and $\Pe$, it is perhaps unsurprising that it is more challenging to determine an analytical expression for the location of the right-hand boundary between two peaks and one peak in figure \ref{fig:Secondary_peak_Regime_diagram}), and unfortunately we have been unable to do so. However, it is relatively straightforward to illustrate why the transition occurs by considering a specific example. 


\begin{figure}
\centering \scalebox{0.55}{\epsfig{file=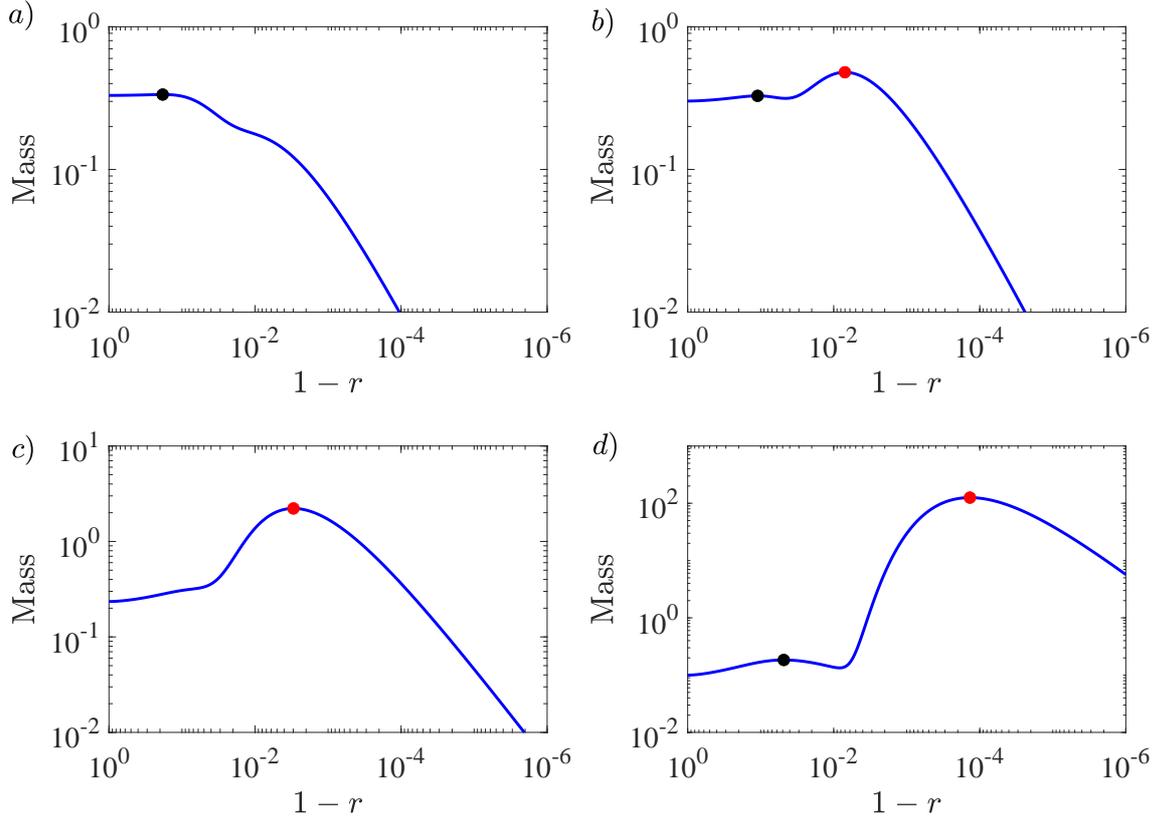}}
\caption{Solute profiles for an evaporating droplet with $\Pe = 10^{2}$ and $\Bo = 10^{3}$ displayed on a doubly-logarithmic plot at $5\%$ ($a)$), $20\%$ ($b)$), $50\%$ ($c)$) and $90\%$ ($d)$) of the drying time. In each figure, the primary peak is indicated by a red circle, while the secondary peak is indicated by a black circle when either exists.}
\label{fig:Secondary_Peak_subsumed} 
\end{figure}

In figure \ref{fig:Secondary_Peak_subsumed}, we plot solute mass profiles for $\Pe = 10^{2}$ and $\Bo = 10^{3}$ at $5\%$, $20\%$, $50\%$ and $90\%$ of the drying time indicating the primary and secondary peaks by red and black circles where appropriate. Note that, for such a large Bond number, the critical time at which we would expect a secondary peak to be present may be found from (\ref{eqn:Critical_tc}) to be $t_{c} \approx 2.8\times10^{-10}$. We see in figure \ref{fig:Secondary_Peak_subsumed}a that, indeed, after $5\%$ of the drying time, the secondary peak has emerged and is visible close to the droplet centre --- moreover, at this stage, the primary peak associated with the coffee ring has yet to fully develop (so that the `one peak' at this stage in figure \ref{fig:Secondary_peak_Regime_diagram}a is in fact the secondary peak!). However, by the time we reach $20\%$ of the drying time, both peaks are clearly visible, with the primary peak now approximately $50\%$ larger than the secondary peak. 

Increasing time further, we see that the primary peak continues to grow rapidly so that, by $50\%$ of the drying time, it is so large, that it has \tit{subsumed} the secondary peak into its upstream tail. That is, the secondary peak is still present according to the $\Pe = \infty$ theory, but due to the fact that $\Pe$ is actually finite and the corresponding presence of the classical coffee ring, we do not see the secondary peak. 

If we then increase $t$ even further, we see that by $90\%$ of the drying time, the secondary peak has \tit{reemerged} from the lee of the primary peak. By this stage of the evaporation process, the primary peak has moved significantly closer to the contact line --- here $1-r_{\mathrm{peak},I}\approx1.4\times10^{-4}$, while the secondary peak is located at $1-r \approx 4.8\times10^{-2}$, so that it is sufficiently far behind the primary peak to be visible.

Thus, the loss of the secondary peak appears to be intrinsically tied to both the location, size and shape of the primary peak. Given that this behaviour largely occurs in the regime in which $\Bo \gg 1$, these properties of the primary peak are given by (\ref{eqn:Primary_Peak_Location}), (\ref{eqn:Primary_Peak_Height}) and the derivative of (\ref{eqn:IMV_inner_LO}), respectively. Clearly, therefore, the behaviour is strongly dependent on $t$, $\Bo$ and $\Pe$ (cf. figure \ref{fig:Primary_Ring_Properties}, for example).  

\subsubsection{Properties of the secondary peak}

Given its dependence on the various parameters of the model, discerning the properties of the secondary peak analytically is challenging, particularly in the $\Bo = O(1)$-regime since, in this case, the peak tends to be situated in the droplet bulk, so that we are unable to use the simpler forms of the inner solution described in \textsection \ref{sec:Moderate_inner}.

\begin{figure}
\centering \scalebox{0.55}{\epsfig{file=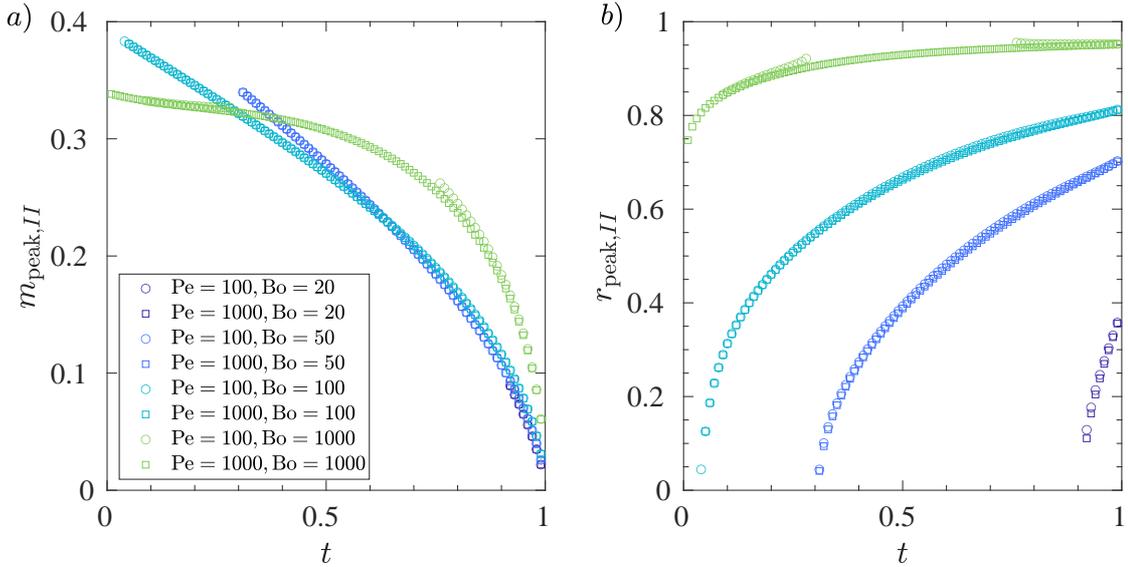}}
\caption{Numerical predictions of ($a)$) the height of the secondary peak, $m_{\mathrm{peak},II}(t)$ and ($b)$) its location $r_{\mathrm{peak},II}(t)$ for different values of $\Pe$, $\Bo$. The symbols denote different P\'{e}clet numbers: $\Pe = 100$ (circles), $\Pe = 1000$ (squares); while the colours denote different Bond numbers: $\Bo = 20$ (purple), $\Bo = 50$ (dark blue), $\Bo = 100$ (light blue), $\Bo = 1000$ (green).}
\label{fig:Secondary_peak_properties} 
\end{figure}

Hence, we utilize the numerical results to track the height $m_{\mathrm{peak},II}(t)$ and location $r_{\mathrm{peak},II}(t)$ of the secondary peak when it exists and we display the results for several different values of $\Pe$, $\Bo$ in figure \ref{fig:Secondary_peak_properties}. In the figure, results for $\Pe = 10^2$ and $\Pe = 10^3$ are denoted by circles and squares, respectively. The Bond number is represented by the colour, with results for $\Bo = 20$ (purple), $50$ (dark blue), $100$ (light blue) and $1000$ (green). It is evident that for each of the Bond numbers represented, increasing the P\'{e}clet number appears to have negligible effect on both the size and location of the secondary peak. However, both properties do vary with the Bond number. In particular, as the Bond number increases, the secondary peak is situated closer to the contact line at the same stage of the drying process, and similarly, for a fixed Bond number, the peak gets closer to the contact line as the droplet evaporates. On the other hand, variations of the secondary peak height with $\Bo$ are much less trivial, although for all of the displayed results, we see that the height of the secondary peak \tit{decreases} as the droplet evaporates. This is in stark contrast to the primary peak, which always grows as more solute is transported to the contact line. 

Thus, we conclude that the secondary peak is predominantly driven by the Bond number. Indeed, it is only for sufficiently large Bond numbers that we find a second peak at all, and the properties of that peak then depend strongly on the size of $\Bo$. The only role played by the P\'{e}clet number appears to be in the disappearance of the secondary peak when it is subsumed by the primary peak, which is typically orders of magnitude larger and always closer to the contact line.  


\section{Summary and discussion}
\label{sec:Summary}

In this paper, we have performed a detailed asymptotic and numerical analysis into the effect of gravity on the famous coffee ring phenomenon observed in solute-laden droplets. In the physically-relevant limit of small droplet capillary number, $\Ca\ll1$, and large solutal P\'{e}clet number, $\Pe\gg1$, we identified two asymptotic regimes based on the size of the Bond number, $\Bo$: 
\begin{enumerate}
\item[i)] a moderate Bond number regime, where $\Bo = O(1)$;
\item[ii)] a large Bond number regime, $\Bo \gg 1$.
\end{enumerate}


In the first of these regimes, gravity acts to flatten the droplet profile from the spherical cap of the zero-gravity problem, while reducing the liquid velocity. Moreover, the asymptotic structure of the solute transport follows exactly that presented by \cite{Moore2021a} for surface tension-dominated droplets, with advection dominating in the droplet bulk, while the competition between advection and diffusion in a boundary layer of width of $O(\Pe^{-2})$ near the pinned contact line drives the nascent coffee ring. Gravity acts to modify the surface tension-dominated solution both through the accumulated mass flux of solute into the contact line and a parameter dependent on the local contact angle. In particular, as the Bond number increases, the height of the nascent coffee ring is reduced --- which is consistent with the reduced flow velocity as $\Bo$ is increased. Moreover, the peak is situated further from the contact line.

To categorize the role of gravity more explicitly, we derived an approximate similarity profile, $\hat{m}_{0}$, for the nascent coffee ring profile, given by
\begin{linenomath}
\begin{equation}
 \frac{\hat{m}_{0}(R,t)}{\Pe_{t}^{2}\mathcal{N}(t;\Bo)} = \frac{2\chi}{3\psi(\Bo)}f\left(\sqrt{R},3,\frac{4\chi}{\psi(\Bo)}\right), \quad R = \Pe_{t}^{2}(1-r)
\end{equation}
\end{linenomath}
where $\Pe_{t} = \Pe/(1-t)$ is the time-dependent P\'{e}clet number, $\mathcal{N}(t;\Bo)$ is the accumulated mass flux of solute at the contact line from the droplet bulk, $\chi$ is the coefficient of the inverse square root singularity in the evaporative flux at the contact line; $\psi(\Bo)$ is the leading order initial  local contact angle; and $f(x,k,l) = l^kx^{k-1}\mbox{e}^{-lx}/\Gamma(k)!$ is the probability density function of a gamma distribution. Clearly, the Bond number acts to scale the coffee ring profile through the accumulated mass flux, while it acts to change the \tit{shape} of the profile through the initial contact angle $\psi(\Bo)$.


In the second regime, the Bond number is large, so that the droplet is approximately flat, with surface tension confined to a narrow region near the pinned contact line --- a `pancake' or `puddle' droplet. Thus, the asymptotic analysis discussed above is no longer valid, since there are two competing boundary layers near the edge of the droplet --- the diffusion boundary layer in the solute transport and the surface tension boundary layer in the droplet free surface profile (and, hence, the liquid velocity). We derived the resulting solute distribution in the most general regime in which the two boundary layers are comparable, which reduces to the assumption that $\alpha = \Bo^{-1/2}\Pe^{2/3} = O(1)$. Under this assumption, diffusion and advection balance in a region of size $\Pe^{-2/3}$ near the contact line, noticeably larger than in the moderate gravity regime. This is a further indication of gravity acting to shift the coffee ring further from the contact line and, moreover, tends to cause shallower solute profiles in the boundary layer region. 

The asymptotic analysis in the large-Bond number regime is more challenging than that in the moderate Bond number regime and, in particular, the nascent coffee ring no longer collapses onto an approximate similarity profile. However, we were able to derive expressions for the location (\ref{eqn:Primary_Peak_Location}) and height (\ref{eqn:Primary_Peak_Height}) of the peak, demonstrating that it still strongly depends on the accumulated mass flux of solute into the contact line alongside the parameter $\alpha$. In particular, increasing $\alpha$ leads to higher coffee rings that are located closer to the contact line.

In each regime, we demonstrated that our asymptotic predictions were in excellent agreement with numerical simulations of the full advection-diffusion problem for the solute mass distribution.  


Alongside the anticipated nascent coffee ring driven by the competition between advection and diffusion of the solute, our asymptotic and numerical analysis also revealed a novel phenomenon: that the solute profile may have a \tit{secondary} peak. The secondary peak was characterized by being situated upstream of and significantly smaller than the primary coffee ring. Moreover, the presence of this peak strongly depended on the Bond number, P\'{e}clet number and evaporation time. 

Further investigation revealed that, for a fixed P\'{e}clet number, there exists a band in $(\Bo,t)$-space at which two peaks are present in the profile. We demonstrated that the onset of this band is independent of the P\'{e}clet number and is caused by the critical point at the centre of the droplet changing in type from a maximum (as in the spherical cap droplet in the $\Bo = 0$ regime) to a minimum. When the critical point at the droplet centre changes type, an internal maximum forms downstream of the centre and it is this that corresponds to the secondary peak. This behaviour only occurs above a critical Bond number, $\Bo_{c} \approx 15.21$, and then only after a given drying time, given by
\begin{linenomath}
\begin{equation}
t_{c}(\Bo) = 1 - \left(\frac{2\mathcal{U}_{1}(\Bo)\mathcal{H}_{0}(\Bo)}{2\mathcal{U}_{1}(\Bo)\mathcal{H}_{0}(\Bo)+\mathcal{H}_{1}(\Bo)\mathcal{U}_{0}(\Bo)}\right)^{2/\mathcal{U}_{0}(\footnotesize{\Bo})}.
\end{equation}
\end{linenomath}
In particular, as $\Bo$ increases, $t_c$ decreases, so the secondary peak emerges earlier in the evaporative process. These predictions were shown to be in excellent agreement to the numerical results and, remarkably, are independent of the P\'{e}clet number.


However, the above analysis suggests that for all $\Bo > \Bo_c$ and $t>t_c$ a secondary peak exists --- something that we did not find in our analysis. The reason for this discrepancy was shown to be due to the presence of the primary peak. In particular, as time increases, the secondary peak is located further from the droplet centre so that it may get subsumed in the tail of the primary peak. For a fixed Bond number, this possibility was shown to depend strongly on both the P\'{e}clet number and the evaporation time; this is due to the fact that the size of the primary peak increases with both $t$ and $\Pe$, while the size of the secondary peak only varies with $t$.

Beyond this subsuming effect, however, we were able to demonstrate that the P\'{e}clet number plays negligible role in the size and location of the secondary peak for a range of Bond numbers, suggesting that this feature may be reliably controlled simply by altering $\Bo$. 


In previous studies of coffee ring formation (e.g. \cite{Deegan2000, popov2005evaporative, Moore2021a}, gravity has frequently been neglected under the assumption of small Bond number, which is a reasonable assumption for sufficiently small droplets. However, given that the Bond number may be increased in an experimental or industrial setting by steadily increasing the droplet radius, the influence of gravity may be of fundamental interest in applications that utilize droplet drying to adaptively control the shape of the residual deposit, such as  colloidal patterning \citep{Harris2007, Choi2010} and fabrication techniques using inkjet printing \citep{Layani2009}.  Our analysis thus plays a dual role in the field. First, we have presented the first formal categorization of the role of gravity in the early-stages of coffee ring formation and given a quantitative prediction of the resulting features of the solute profile. Second, we have found a novel phenomenon --- the secondary peak --- which may also be exploited in such processes, particularly when the size of the primary peak can be carefully controlled. This is particularly relevant given that the secondary peak emerges at a relatively moderate critical Bond number. 

There are, naturally, limitations to our analysis. Throughout, we have assumed that the contact line remains pinned as the droplet evaporates. This has been shown to be a reasonable assumption for many configurations (see, for example, the experiments in \cite{Deegan2000, Kajiya2008, howard2023surfactant}) and may further be enhanced by solute aggregation near the edge of the droplet (\cite{Orejon2011, Weon2013,Larson2014}). However, at late stages of the evaporation, the contact line may depin and become mobile, moving inwards towards the droplet centre. The contact line may then become pinned at a new location and the process may repeat. This behaviour is known as `stick-slip' evaporation and represents an important class within the field that is beyond the scope of the present study, but may represent an interesting future direction in terms of the effect of gravity, particularly with the presence of the secondary peak and its associated increased solute mass, which may promote re-pinning.

Another effect that we have neglected in the present analysis is the possibility of solute becoming trapped at the free surface of the droplet. If this occurs, the solute is then transported to the contact line along the free surface, and has been suggested as an alternative mechanism for coffee ring formation (\cite{Kang2016}). This behaviour has been demonstrated to occur for a wide variety of droplets but is more pronounced for droplets with large contact angles \cite{Kang2016} or when vertical diffusion happens over a longer timescale than evaporation \cite{d2022evolution}. Since we deal with the opposite case of a thin drop with fast vertical diffusion (i.e. so that the solute concentration may be assumed to be uniform across the droplet to leading order), we have not considered this phenomenon here. It would be interesting to see how such behaviour impacted the solute profile in the current case, although it should be noted that the aforementioned studies neglect gravity entirely.

A further aspect that would form the basis of an exciting future study surrounds the assumption that the solute is dilute in the droplet. Naturally, the build up of the solute in the coffee ring means that the concentration rapidly approaches levels where finite particle size effects can no longer be ignored. This has been analysed in detail for surface tension-dominated droplets in \cite{Moore2021a, Moore2022} and a similar analysis would follow here with the inclusion of gravity. One possible aspect that would differentiate droplets where gravity is included is in the vicinity of the secondary peak. It is an interesting open question as to whether the dilute assumption may also break down in the vicinity of the secondary peak as well as the primary. Once finite particle size effects become important, there are a number of different approaches that can be followed to continue the analysis, such as a sharp transition between a dilute and jammed region (\cite{popov2005evaporative}), using a viscosity and solute diffusivity that vary with concentration (\cite{kaplan2015evaporation}) or through more complicated two-phase suspension models (see, for example, \cite{Guazzelli2018}). 

Our analysis has concentrated on a diffusive evaporative model, while there are many situations where other evaporative models may be appropriate. Examples include water evaporating on glass, which may more appropriately be modelled using a kinetic evaporative model (\cite{Murisic2011}), droplets evaporating above a bath of the same liquid, where the evaporation is effectively constant (\cite{boulogne2016coffee}) and situations where a mask is used to control the evaporative flux so that it is stronger towards the centre (\cite{vodolazskaya2017modeling}). The analysis herein could readily be extended to such situations, although we note that for evaporative fluxes with different --- including non-singular --- behaviour close to the contact line, the size of the boundary layer regions near the contact line in which solutal diffusion and surface tension are relevant will change accordingly, as for surface tension-dominated droplets \citep{Moore2021a}.

A future direction of interest would be to extend the analysis herein to non-axisymmetric droplets. Such droplets occur widely in applications, particularly in printing OLED/AMOLED screens (see, for example, \cite{mai200753,huo2020real}). It is well-known that droplet geometry plays a strong role in the behaviour of the evaporative flux (\cite{Saenz2017, Wray2023}) and the transient and final deposit profiles (\cite{FreedBrown2015,Saenz2017,Moore2022}). It would be of significant theoretical and practical interest to explore the behaviour of the secondary peak in such problems as well.

Finally, we note that another context in which gravity may play an important role is that of binary/multi-component droplets, particularly in situations where the different fluids have different densities. Multi-component droplets occur widely, from commercial alcohols such as whiskey and ouzo (\cite{tan2019porous, carrithers2020multiscale}) to various inks (\cite{shargaieva2020hybrid}). While it would be certainly of interest to extend the analysis presente here to such droplets, a careful treatment of the internal flow would be needed, as the multi-component nature of the droplet significantly complicates the dynamics (\cite{li2019gravitational}).

\textbf{Acknowledgments} MRM would like to acknowledge the support of EPSRC grant EP/X035646/1.

\tbf{Declaration of Interests.} The authors report no conflict of interests.


\appendix 

\section{Matyched asymptotic analysis in the limit of large $\Bo$, $\alpha = O(1)$}
\label{appendix:Large_Bo}

In this appendix, we present the asymptotic solution of the solute transport problem in the limit in which $\Bo, \Pe\gg1$ and
\begin{linenomath}
\begin{equation}
\alpha = \Bo^{-1/2}\Pe^{2/3} = O(1).
\end{equation}
\end{linenomath}
For convenience, we choose to use $\Pe^{-2/3}$ as our small parameter in the asymptotic expansions. Moreover, it transpires that it is easier to analyse the integrated mass variable formulation of the problem  (\ref{eqn:IMV_AdvDiff})--(\ref{eqn:SoluteMass}).

\subsection{Outer region}
\label{sec:Outer}

In the droplet bulk, $1-r$ is $O(1)$, and we recall from (\ref{eqn:h0}) that  the droplet free surface $h$ is flat to all orders and that the velocity is given by (\ref{eqn:u0}). 
Upon substituting these expressions into (\ref{eqn:IMV_AdvDiff}) and (\ref{eqn:IMV_IC}), and then expanding $\mathcal{M}(r,t) = \mathcal{M}_{0}(r,t) + \Pe^{-2/3}\mathcal{M}_{1}(r,t) + O(\Pe^{-4/3})$ as $\Pe\rightarrow\infty$, we find to leading-order
\refstepcounter{equation}
\begin{linenomath}
$$
 \frac{\partial\mathcal{M}_{0}}{\partial t} + \frac{u_{0}}{4}\frac{\partial\mathcal{M}_{0}}{\partial r}= 0 \quad \mbox{for} \quad 0<r,t<1, \quad \mathcal{M}_{0}(r,0) = \frac{r^{2}}{2\pi}\quad \mbox{for} \quad 0<r<1. 
 \eqno{(\theequation{\mathit{a},\mathit{b}})}\label{eqn:Leading_order_outer_1}
$$
\end{linenomath}
This may be solved using the method of characteristics, yielding
\begin{linenomath}\begin{equation}
 \mathcal{M}_{0}(r,t) = \frac{(1-t)r^{2}}{2\pi} + \frac{\sqrt{1-t}(1-\sqrt{1-t})}{\pi}(1-\sqrt{1-r^{2}}).
 \label{eqn:IMV_outer}
\end{equation}\end{linenomath}
We see that this solution automatically satisfies the boundary condition (\ref{eqn:IMV_BC}a). 

At $O(\Pe^{-2/3})$, the problem for $\mathcal{M}_{1}(r,t)$ is given by
\refstepcounter{equation}
\begin{linenomath}
$$
 \frac{\partial\mathcal{M}_{1}}{\partial t} + \frac{u_{0}}{4}\frac{\partial\mathcal{M}_{1}}{\partial r}= -\frac{\alpha u_{1}}{4}\frac{\partial\mathcal{M}_{0}}{\partial r} \quad \mbox{for} \quad 0<r,t<1, \quad \mathcal{M}_{1}(r,0) = 0 \quad \mbox{for} \quad 0<r<1. 
 \eqno{(\theequation{\mathit{a},\mathit{b}})}\label{eqn:Second_order_outer_1}
$$
\end{linenomath}
for $0<r<1$, $0<t<1$, while the initial condition is given by $\mathcal{M}_{1}(r,0) = \alpha r^{2}/\pi$ for $0<r<1$. This may be solved in a similar manner using the method of characteristics, yielding 
\begin{linenomath}\begin{equation}
\begin{aligned}
 \mathcal{M}_{1}(r,t) = & \; \frac{2\alpha\kappa(r,t)}{\pi}(1-\kappa(r,t))\log\left(\frac{\sqrt{1-t}-\kappa(r,t)}{1-\kappa(r,t)}\right) + \frac{\alpha}{\pi}(1-(1-\kappa(r,t))^{2}),
 \end{aligned}
 \label{eqn:IMV_outer_SO}
\end{equation}\end{linenomath}
where $\kappa(r,t) = \sqrt{1-t}(1-\sqrt{1-r^{2}})$.

Expanding the leading-order solution (\ref{eqn:IMV_outer}) as we approach the contact line, we have 
\begin{linenomath}\begin{equation}
 \mathcal{M}_{0}(r,t) \sim \frac{\sqrt{1-t}}{\pi} - \frac{(1-t)}{2\pi} - \frac{\sqrt{2(1-t)}}{\pi}(1-\sqrt{1-t})\sqrt{1-r} - \frac{(1-t)}{\pi}(1-r) + O((1-r)^{3/2})
 \label{eqn:IMV_local}
\end{equation}\end{linenomath}
as $r\rightarrow1^-$. Notably, this means that the leading-order outer solute mass $m_{0}$ is singular at the contact line, which gives a strong indication of the importance of diffusive effects local to the edge of the droplet. This is in stark contrast to the $\Bo = O(1)$ solution, where the outer solute mass was square root \tit{bounded} as $r\rightarrow1^-$.

A similar expansion of (\ref{eqn:IMV_outer_SO}), yields
\begin{linenomath}\begin{equation}
\begin{aligned}
 \mathcal{M}_{1}(r,t) \sim & \; \frac{\alpha\sqrt{1-t}(1-\sqrt{1-t})}{\pi}\log(1-r) +  \frac{\alpha\sqrt{1-t}(1-\sqrt{1-t})}{\pi}\log\left(\frac{2(1-t)}{(1-\sqrt{1-t})^{2}}\right) + \\
 & \; \frac{\alpha}{\pi}(1-(1-\sqrt{1-t})^{2}) + O(\sqrt{1-r}\log(1-r))
 \label{eqn:IMV_local_SO}
\end{aligned}
\end{equation}\end{linenomath}
as $r\rightarrow1^{-}$.
We can clearly see this will necessitate an inner expansion that contains logarithmic terms; similar behaviour is displayed for surface tension-dominated drops under different evaporative fluxes \citep{Moore2021a}.

Finally, if we expand the solute mass $m \sim m_{0}$ as $\Pe\rightarrow0$ in  (\ref{eqn:SoluteMass}), we find
\begin{linenomath}\begin{equation}
 m_{0}(r,t) = \frac{\sqrt{1-t}}{\pi\sqrt{1-r^{2}}}\left[1-\sqrt{1-t}(1-\sqrt{1-r^{2}})\right].
 \label{eqn:Mass_outer}
\end{equation}\end{linenomath}
Whilst we could proceed to $O(\Pe^{-2/3})$ in the solute mass expansion in the outer region, we shall not require this when constructing a composite profile that is valid to $O(1)$ throughout the droplet, so we do not present this here.

\subsection{Inner region}
\label{sec:Inner}

Recalling (\ref{eqn:h_local_Bo_large})--(\ref{eqn:u_local_Bo_large}), (\ref{eqn:Local_balance}) and (\ref{eqn:alpha}), in order to retain a balance between the advective and diffusive effects in (\ref{eqn:IMV_AdvDiff}) close to the contact line, we set
\begin{linenomath}\begin{equation}
 r = 1 - \Pe^{-2/3} \tr, \quad u = \Pe^{-1/3} \tu, \quad h = \th, \quad \mathcal{M} = \tM, \quad m = \Pe^{2/3}\tilde{m} 
 \label{eqn:Scalings_Pe}
\end{equation}\end{linenomath}
in (\ref{eqn:IMV_AdvDiff})--(\ref{eqn:SoluteMass}). Note that we therefore have
\begin{linenomath}\begin{equation}
 \th = \th_0 + \Pe^{-2/3}\th_1 + O(\Pe^{-4/3}), \quad \tu = \tu_0 + \Pe^{-1/3}\tu_1 + \Pe^{-2/3}\tu_2 + O(\Pe^{-1}) 
\end{equation}\end{linenomath}
as $\Pe\rightarrow\infty$ where
\begin{linenomath}\begin{equation}
 \th_0(\tr,t) = \bar{h}_0(\tr/\alpha,t), \quad \th_1(\tr,t) = \alpha\bar{h}_1(\tr/\alpha,t), 
\end{equation}\end{linenomath}
and $\bar{h}_{0}$, $\bar{h}_{1}$ are given by (\ref{eqn:h0_inner})--(\ref{eqn:h1_inner}), and
\begin{linenomath}\begin{equation}
 \tu_0(\tr,t) = \sqrt{\alpha}\bar{u}_{0}(\tr/\alpha,t), \quad \tu_1(\tr,t) =  \alpha\bar{u}_{1}(\tr/\alpha,t), \quad \tu_2(\tr,t) = \alpha^{3/2}\bar{u_{2}}(\tr/\alpha,t)
\end{equation}\end{linenomath}
and $\bar{u}_{0}$, $\bar{u}_{1}$, $\bar{u}_{2}$ are given by (\ref{eqn:u0_inner}))--(\ref{eqn:u2_inner}).

Seeking an asymptotic expansion of the integrated mass of the form $\tM = \tM_{0} + \Pe^{-1/3}\tM_{1} + \Pe^{-2/3}\log\Pe^{-2/3}\tM_{2} + \Pe^{-2/3}\tM_{3} + o(\Pe^{-2/3})$ as $\Pe\rightarrow\infty$, we find that the leading-order inner problem is given by
\begin{linenomath}\begin{equation}
 \frac{\partial^{2}\tM_{0}}{\partial \tr^{2}} + \left(\tu_{0} - \frac{1}{\th_0}\frac{\partial \th_0}{\partial \tr}\right)\frac{\partial\tM_{0}}{\partial r} = 0, \quad \mbox{for} \quad \tr>0, \; 0<t<1, 
  \label{eqn:IMV_inner_LO_problem}
\end{equation}\end{linenomath}
subject to the boundary condition $\tM_{0}(0,t) = 1/2\pi$ for $0<t<1$ and, in order to match with the local expansion of leading-order-outer solution at the contact line (\ref{eqn:IMV_local}), we must have
\begin{linenomath}\begin{equation}
 \tM_{0}\rightarrow \frac{\sqrt{1-t}}{\pi} - \frac{(1-t)}{2\pi} \quad \mbox{as} \quad \tr\rightarrow\infty,
 \label{eqn:IMV_inner_LO_problem_2}
\end{equation}\end{linenomath}
Defining the integrating factor
\begin{linenomath}\begin{equation}
\begin{aligned}
 I(\tr,t) = & \;  \left(\frac{1}{1-\mbox{e}^{-\tr/\alpha}}\right)\mbox{exp}\left(\frac{2\sqrt{2}}{(1-t)}\int_{0}^{\tr}\frac{\sqrt{\xi}}{1-\mbox{e}^{-\xi/\alpha}}\,\mbox{d}\xi\right),
 \label{eqn:Integrating_Factor}
 \end{aligned}
\end{equation}\end{linenomath}
we find that the solution is given by
\begin{linenomath}\begin{equation}
 \tM_{0}(\tr,t) = \frac{1}{2\pi} + B_{0}(t)\int_{0}^{\tr}\frac{1}{I(s,t)}\,\mbox{d}s,
 \label{eqn:IMV_inner_LO}
\end{equation}\end{linenomath}
where
\begin{linenomath}\begin{equation}
 B_{0}(t) = -\frac{1}{\pi}\left(1-\sqrt{1-t}-\frac{t}{2}\right)\left(\int_{0}^{\infty}\frac{1}{I(s,t)}\,\mbox{d}s\right)^{-1}. 
 \label{eqn:B0}
\end{equation}\end{linenomath}
We note here that the first term on the right-hand side of $B_{0}(t)$ is simply the leading-order accumulated mass at the contact line as a function of time, $\mathcal{N}(t)$, that is
\begin{linenomath}\begin{equation}
 \mathcal{N}(t) = \frac{1}{4}\int_{0}^{t} (m_{0} u_{0})(1^{-},\tau)\,\mbox{d}\tau = \frac{1}{\pi}\left(1 - \sqrt{1-t} - \frac{t}{2}\right).
 \label{eqn:MassFluxIn}
\end{equation}\end{linenomath}
It is worth noting the similarities between (\ref{eqn:MassFluxIn}) and the equivalent expression for a surface-tension dominated drop evaporating under a constant evaporative flux \citep{FreedBrown2015,Moore2021a}.

At $O(\Pe^{-1/3})$, we have
\begin{linenomath}\begin{equation}
 \frac{\partial^{2}\tM_{1}}{\partial \tr^{2}} + \left(\tu_{0} - \frac{1}{\th_0}\frac{\partial \th_0}{\partial \tr}\right)\frac{\partial\tM_{1}}{\partial r} = 4\frac{\partial\tM_{0}}{\partial t} - \tu_{1}\frac{\partial\tM_{0}}{\partial\tr} \quad \mbox{for} \quad \tr>0,\;0<t<1,
 \label{eqn:IMV_inner_SO_problem}
 \end{equation}\end{linenomath}
 subject to $\tM_{1}(0,t) = 0$ for $0<t<1$ and the far-field matching condition
 \begin{linenomath}\begin{equation}
 \tM_{1}\rightarrow -\frac{\sqrt{2(1-t)}}{\pi}(1-\sqrt{1-t})\sqrt{\tr} \quad \mbox{as} \quad \tr\rightarrow\infty.
 \label{eqn:IMV_inner_SO_problem_2}
\end{equation}\end{linenomath}
While in practice it may be easier to find $\tM_{1}(\tr,t)$ from (\ref{eqn:IMV_inner_SO_problem})--(\ref{eqn:IMV_inner_SO_problem_2}) numerically, for posterity, we state that this boundary value problem has solution
\begin{linenomath}\begin{equation}
\begin{aligned}
 \tM_{1}(\tr,t) = & \; \int_{0}^{\tr}\frac{1}{I(s,t)} \left(\int_{0}^{s}\left(4\frac{\partial\tM_{0}}{\partial t} - \tu_{1}\frac{\partial\tM_{0}}{\partial \tr}\right)I(\sigma,t)\,\mbox{d}\sigma\right)\,\mbox{d}s + B_{1}(t)\int_{0}^{\tr}\frac{1}{I(s,t)}\,\mbox{d}s,
 \end{aligned}
 \label{eqn:IMV_inner_SO}
\end{equation}\end{linenomath}
where
\begin{linenomath}\begin{equation}
\begin{aligned}
 B_{1}(t) = & \; -\left(\int_{0}^{\infty}\left\{\frac{1}{I(s,t)} \left(\int_{0}^{s}\left(\frac{4\partial\tM_{0}}{\partial t} - \tu_{1}\frac{\partial\tM_{0}}{\partial \tr}\right)I(\sigma,t)\,\mbox{d}\sigma\right)- \sqrt{2}(1-t)\frac{\partial\tM_{0}}{\partial t}\frac{1}{\sqrt{s}}\right\}\,\mbox{d}s \right)\left(\int_{0}^{\infty}\frac{1}{I(s,t)}\,\mbox{d}s\right)^{-1},
 \label{eqn:B1}
 \end{aligned}
\end{equation}\end{linenomath}
is chosen to kill the $O(1)$-term in the far-field expansion of $\tM_{1}(\tr,t)$.

The $O(\Pe^{-2/3}\log\Pe^{-2/3})$-problem is given by
\begin{linenomath}\begin{equation}
 \frac{\partial^{2}\tM_{2}}{\partial \tr^{2}} + \left(\tu_{0} - \frac{1}{\th_0}\frac{\partial \th_0}{\partial \tr}\right)\frac{\partial\tM_{2}}{\partial r} = 0 \quad \mbox{for} \quad \tr>0, \; 0<t<1,
\end{equation}\end{linenomath}
subject to $\tM_{2}(0,t) = 0$ for $0<t<1$ and the far-field matching condition
\begin{linenomath}\begin{equation}
 \tM_{2}\rightarrow \frac{\alpha\sqrt{1-t}(1-\sqrt{1-t})}{\pi} \quad \mbox{as} \quad \tr\rightarrow\infty.
\end{equation}\end{linenomath}
The solution may be found in a similar manner to the leading-order problem, yielding
\begin{linenomath}\begin{equation}
 \tM_{2}(\tr,t) = B_{2}(t)\int_{0}^{\tr}\frac{1}{I(s,t)}\,\mbox{d}s,
 \label{eqn:IMV_inner_TO}
\end{equation}\end{linenomath}
where
\begin{linenomath}\begin{equation}
 B_{2}(t) = \frac{\alpha\sqrt{1-t}(1-\sqrt{1-t})}{\pi}\left(\int_{0}^{\infty}\frac{1}{I(s,t)}\,\mbox{d}s\right)^{-1}. 
 \label{eqn:B2}
\end{equation}\end{linenomath}

Lastly, at $O(\Pe^{-2/3})$, we have
\begin{linenomath}\begin{equation}
 \frac{\partial^{2}\tM_{3}}{\partial \tr^{2}} + \left(\tu_{0} - \frac{1}{\th_{0}}\frac{\partial \th_{0}}{\partial \tr}\right)\frac{\partial\tM_{3}}{\partial r}  =  4\frac{\partial\tM_{1}}{\partial t} - \tu_{1}\frac{\partial\tM_{1}}{\partial\tr} - \tu_{2}\frac{\partial\tM_{0}}{\partial\tr} - \frac{1}{\th_{0}}\left(\frac{\th_{1}}{\th_{0}}\frac{\partial\th_{0}}{\partial\tr} - \frac{\partial\th_{1}}{\partial\tr}\right)\frac{\partial\tM_{0}}{\partial\tr} - \frac{\partial\tM_{0}}{\partial\tr}=: \mathcal{V}(\tr,t)
 \label{eqn:IMV_inner_FO_problem}
\end{equation}\end{linenomath}
for $\tr>0$, $0<t<1$, subject to $\tM_{3}(0,t) = 0$ for $0<t<1$ and the far-field condition
\begin{linenomath}\begin{equation}
\tM_{3}\rightarrow - \frac{(1-t)}{\pi}\tr +\left(\frac{\alpha\sqrt{1-t}(1-\sqrt{1-t})}{\pi}\right)\left(\log{\tr} + \log\left(\frac{2(1-t)}{(1-\sqrt{1-t})^{2}}\right)\right) + \frac{\alpha}{\pi}(1-(1-\sqrt{1-t})^{2}) \quad \mbox{as} \quad \tr\rightarrow\infty. 
\end{equation}\end{linenomath}
The solution is given by
\begin{linenomath}\begin{equation}
\begin{aligned}
 \tM_{3}(\tr,t) = & \; \int_{0}^{\tr}\frac{1}{I(s,t)} \left(\int_{0}^{s} \mathcal{V}(\sigma,t)I(\sigma,t)\,\mbox{d}\sigma\right)\,\mbox{d}s + B_{3}(t)\int_{0}^{\tr}\frac{1}{I(s,t)}\,\mbox{d}s,
 \end{aligned}
 \label{eqn:IMV_inner_FO}
\end{equation}\end{linenomath}
where
\begin{linenomath}\begin{equation}
\begin{aligned}
 B_{3}(t) = & \; \left[- \int_{1}^{\infty}\left\{\frac{1}{I(s,t)} \left(\int_{0}^{s}\mathcal{V}(\sigma,t)I(\sigma,t)\,\mbox{d}\sigma\right) + \frac{(1-t)}{\pi} - \frac{\alpha\sqrt{1-t}(1-\sqrt{1-t})}{\pi s}\right\}\,\mbox{d}s \right. \\
 & \; \left. -\int_{0}^{1}\frac{1}{I(s,t)} \left(\int_{0}^{s}\mathcal{V}(\sigma,t)I(\sigma,t)\,\mbox{d}\sigma\right)\,\mbox{d}s - \frac{(1-t)}{\pi} + \frac{\alpha}{\pi}(1-(1-\sqrt{1-t})^{2}) \right. \\
 & \; \left. + \frac{\alpha\sqrt{1-t}(1-\sqrt{1-t})}{\pi}\log\left(\frac{2(1-t)}{(1-\sqrt{1-t})^{2}}\right)\right]\left(\int_{0}^{\infty}\frac{1}{I(s,t)}\,\mbox{d}s\right)^{-1},
 \label{eqn:B3}
 \end{aligned}
\end{equation}\end{linenomath}
has been chosen to satisfy the correct far-field behaviour.

We are now in a position to find the inner solution for the solute mass. By substituting the scalings (\ref{eqn:Scalings_Pe}) into (\ref{eqn:SoluteMass}), we see that
\begin{linenomath}\begin{equation}
 \tilde{m} = -\frac{1}{1 - \Pe^{-2/3}\tr}\frac{\partial\tM}{\partial\tr},
\end{equation}\end{linenomath}
so that expanding $\tilde{m} = \tilde{m}_{0} + \Pe^{-1/3}\tilde{m}_{1} + \Pe^{-2/3}\log\Pe^{-2/3}\tilde{m}_{1} +\Pe^{-2/3}\tilde{m}_{2}$ as $\Pe\rightarrow\infty$, we have
\begin{linenomath}\begin{equation}
 \tilde{m}_{0} = -\frac{\partial\tM_{0}}{\partial\tr}, \quad \tilde{m}_{1} = -\frac{\partial\tM_{1}}{\partial\tr}, \quad \tilde{m}_{2} = -\frac{\partial\tM_{2}}{\partial\tr}, \quad \tilde{m}_{3} = -\frac{\partial\tM_{3}}{\partial\tr} - \tr\frac{\partial\tM_{0}}{\partial\tr}.
 \label{eqn:Mass_inner}
\end{equation}\end{linenomath}

\subsection{Composite solutions}
\label{sec:Composite}

We now have all of the necessary components needed to construct (additive) composite solutions for comparison to the numerical results. 

To construct a composite solution for the integrated mass variable, we combine the first two outer solutions (\ref{eqn:IMV_outer}) and (\ref{eqn:IMV_outer_SO}), the first four inner solutions (\ref{eqn:IMV_inner_LO}), (\ref{eqn:IMV_inner_SO}), (\ref{eqn:IMV_inner_TO}) and (\ref{eqn:IMV_inner_FO}), the overlap terms given by (\ref{eqn:IMV_local})--(\ref{eqn:IMV_local_SO}) using Van Dyke's matching rule \cite{VanDyke1964}, which yields
\begin{linenomath}\begin{equation}
\begin{aligned}
 \mathcal{M}_{\mathrm{comp}}(r,t) = & \; \mathcal{M}_{0}(r,t) + \Pe^{-2/3}\mathcal{M}_{1}(r,t) + \tM_{0}\left(\Pe^{2/3}(1-r),t\right) + \Pe^{-1/3}\tM_{1}\left(\Pe^{2/3}(1-r),t\right)  + \\
 & \; \Pe^{-2/3}\log\Pe^{-2/3}\tM_{2}\left(\Pe^{2/3}(1-r),t\right) + \Pe^{-2/3}\tM_{3}\left(\Pe^{2/3}(1-r),t\right) -\\
 & \; \left[\frac{\sqrt{1-t}}{\pi} - \frac{(1-t)}{2\pi} - \frac{\sqrt{2(1-t)}}{\pi}(1-\sqrt{1-t})\sqrt{1-r} - \frac{(1-t)}{\pi}(1-r) + \right. \\
 & \; \left. \Pe^{-2/3}\left(\frac{\alpha}{\pi}(1-(1-\sqrt{1-t})^{2}) + \frac{\alpha\sqrt{1-t}(1-\sqrt{1-t})}{\pi}\left(\log(1-r) + \log\left(\frac{2(1-t)}{(1-\sqrt{1-t})^{2}}\right)\right)\right) \right].
 \label{eqn:IMV_composite}
 \end{aligned}
\end{equation}\end{linenomath}
This composite solution is valid up to and including $O(\Pe^{-2/3})$ throughout the whole of the droplet.

Similarly, for the solute mass, the equivalent composite profile is compiled by taking the first outer solution (\ref{eqn:Mass_outer}) as well as the first four inner solutions given by (\ref{eqn:Mass_inner}), so that, accounting for the overlap contributions,
\begin{linenomath}\begin{equation}
\begin{aligned}
 m_{\mathrm{comp}}(r,t) = &\; m_{0}(r,t) + \Pe^{2/3}\tilde{m}_{0}\left(\Pe^{2/3}(1-r),t\right) + \Pe^{1/3}\tilde{m}_{1}\left(\Pe^{2/3}(1-r),t\right)  + \\
 & \; \log\Pe^{-2/3}\tilde{m}_{2}\left(\Pe^{2/3}(1-r),t\right) + \tilde{m}_{3}\left(\Pe^{2/3}(1-r),t\right)  - \\
 & \; \frac{\sqrt{(1-t)}(1-\sqrt{1-t})}{\sqrt{2}\pi\sqrt{1-r}} -\frac{(1-t)}{\pi} .
 \label{eqn:Mass_composite}
 \end{aligned}
\end{equation}\end{linenomath}
We note that this composite solution is valid up to and including $O(1)$ throughout the droplet.

\section{Numerical solution of the solute transport problem}
\label{appendix:Numerics}

In this section, we outline the numerical scheme for solving the advection-diffusion problem (\ref{eqn:IMV_AdvDiff})--(\ref{eqn:IMV_IC}) for the integrated mass variable $\mathcal{M}(r,t)$. As discussed previously, the integrated mass variable formulation is advantageous when solving numerically, since it is mass-preserving and has simple-to-implement Dirichlet boundary conditions. 

Our numerical method is an adaptation of that discussed in \cite{Moore2021a} for the $\Bo = 0$ regime. We utilize central differences with gridpoints clustered close to the contact line, where there are rapid changes in behaviour associated with the coffee ring. We choose a uniform grid in the variable $\zeta \in[0,1]$, where
\begin{linenomath}
\begin{equation}
r = \frac{1-\ell^\zeta}{1-\ell},
\end{equation}
\end{linenomath}
and $\ell$ is taken to coincide with the smallest of the two boundary layers; that is, $\ell = \kappa(1-t_c)$ where $\kappa = \min\left\{\Bo^{-1/2}, \Pe^{-2/3}\right\}$ and $t_c$ is the final computation time. Note that these boundary layers are in the context of large Bond number; when $\Bo = O(1)$, we have both increased the number of nodes in the discretization and chosen $\ell = \Pe^{-2}$ to ensure we capture the diffusive boundary layer in this regime.

Even when it is present, the secondary peak does not exhibit such extreme behaviour, with a much shallower profile than the primary peak, so provided that the discretization is chosen suitably small, the secondary peak is captured well without special considerations.   
The resulting system is solved using ode15s in MATLAB and incorporates complex step differentiation to compute the Jacobian (\cite{Shampine2007}). The veracity of the simulations has been confirmed with stringent convergent checks alongside the excellent comparisons to the asymptotic results in both the order unity Bond number regime and the large Bond number regime (cf. figures \ref{fig:Composite_Comparisons_Moderate_Bo}, \ref{fig:Composite_Comparisons_Pe3_Bo_1e4}).


\bibliographystyle{jfm}
\nocite{*}


\bibliography{GravityDrops.bib}  

\end{document}